\documentclass{ws-p8-50x6-00}
\usepackage{epsfig}

     \newcommand{\pathnow}{}
\begin{document}\hbadness=10000
\title{The Strange Quark-Gluon Plasma
}
\author{Johann Rafelski \ and \ Giorgio Torrieri
}
\address{Department of Physics, University of Arizona, 
        Tucson, AZ, 85721}
\author{Jean Letessier}
\address{Laboratoire de Physique Th\'eorique et Hautes Energies\\
Universit\'e Paris 7, 2 place Jussieu, F--75251 Cedex 05
}
\maketitle
\begin{center}
\vskip -6.4cm
{
Presented in October 2000 at the 6th International Workshop on\\
{\bf Relativistic Aspects of Nuclear Physics}\\
Tabatinga, Sao Paulo, Brazil}
\vskip 5.3cm
\end{center}
\abstracts{
Strangeness flavor has turned out to be a very effective
diagnostic tool of relativistic heavy ion physics. 
The absolute yield provides information about conditions 
arising  in first instants of the interaction.  
Strange hadrons are abundant allowing a precise study 
of the chemical freeze-out conditions in the dense matter fireball. 
The thermal and chemical freeze-out 
universality seen in strange hadrons confirms sudden
hadronization as the breakup mechanism. A 
plausible cause of sudden fireball breakup is the mechanical instability
arising when a quark-gluon plasma phase  
supercools deeply in its adiabatic expansion.
Applying these ideas, we interpret CERN-SPS and 
the first results on 
hadronic particle ratios obtained at RHIC.}

\section{Introduction}\label{intro}
It is believed today that a new state of matter has been formed 
in relativistic nuclear collisions at CERN. The question
is if  this is the hot quark matter (quark-gluon
plasma) state of matter, which has been for 30 years 
a predicted new form of elementary matter: as early as 1970,
the development of the quark model lead to a first consideration
of quark matter stellar structure.~\cite{Ito70} 

These ideas were deepened by
the development of the  quantum many body theory of 
quark matter,~\cite{Car73,ILL74} which lead on to the
formal recognition within the framework of asymptotically free 
quantum-chromodynamics (QCD) that a very high temperature 
perturbative quark matter state must exist.~\cite{Col75} 
Other reasoning based on a study of the `boiling state' of dense
hadron gas within the scheme of Hagedorn's statistical 
bootstrap of hadronic matter has independently
lead from a different direction
to the consideration of the transition to
a hadron substructure phase.~\cite{Hag84} 
Considering the present day  lattice-QCD numerical 
simulations,~\cite{Kar00} 
we understand today quite well the properties
of the baryon number free 
quark-gluon plasma (QGP), as we today 
call hot quark matter

However, it is still common  to study ad-hoc
ideas about new phases of fundamental matter. 
This is a natural approach considering 
the history of nuclear physics which developed without 
fundamental knowledge about the quark/QCD nature of nuclear 
interactions. Nuclear physics
advanced by the development of  
models, and this tradition remains strong.
In fact there was little initial interest in quarks,
gluons, and QCD among nuclear physisists. The first of the series of formative workshops
in the field of relativistic heavy ion collisions,
the ``Bear Mountain'' meeting   in November 1974
had not a single  mention of quarks, let alone of quark matter.
At  the time  ultra dense nuclei,
multi-hyperon nuclear states,  and
 ``pion condensates'', were all considered as the 
discovery potential of these new and coming tools of 
nuclear physics research.  

As the ideas about QGP formation in high energy nuclear 
collision matured, an unexpected challenge  emerged 20 years ago: how
can the locally deconfined state which exists a mere 10$^{-22}$s
 be distinguished from the gas of confined hadrons? This is 
also a question of principle, considering that both quark 
and hadron pictures of the reaction could be equivalent, for
it has been argued that a quark-gluon based description 
is merely a change of Hilbert space expansion basis, 
if the rules of quantum mechanics with pure 
states are considered. This view contradicts the intuition, 
and about everybody in our field has come to believe that QGP 
phase is observable, since the quantum solution decoheres, 
even if the argument is being made with a lot
of hand waving. 

In the Galilean tradition, such a difficult question 
about existence and
observability of a new phase of elementary matter, the 
QGP, must be settled by an experiment. 
This requires that a probe of QGP operational on the collision
time scale of $5\cdot 10^{-23}$\,s is developed, which is sensitive 
to the local color charge deconfinement, and that it 
depends on the gluon degree of freedom, which is the characteristic 
new ingredient of the quark matter phase.

One of us (JR) proposed strangeness as the signature of 
QGP noticing more than 20 years ago that 
when color bonds are broken, the chemically
equilibrated  deconfined state has an unusually high number of strange
quark pairs.~\cite{abundance} 
A study of the dynamics of strangeness  (chemical) equilibration 
process has shown that only the  gluon component in the QGP is 
capable to produce strangeness rapidly~\cite{RM82}, allowing the 
formation of (nearly) chemically equilibrated dense phase of deconfined, 
hot, strangeness-rich quark matter in relativistic nuclear collisions. 
Therefore, strangeness enhancement is related directly to the
presence of gluons in QGP.

The high density of strangeness available in the fireball 
favors during hadronization the 
formation of multi-strange hadrons,~\cite{Raf82,RD83} which 
are produced quite rarely if only individual hadrons 
collide.~\cite{Koc85,KMR86} A large enhancement 
of multi strange (anti)baryons has been continuously 
advanced and defended by one of us (JR) as a characteristic
signature of the QGP over the past 20 years. 
A systematic enhancement has been now reported, rising with 
strangeness content.~\cite{WA97p} 
Here the enhancement of strange antibaryons is  reported 
with a base being the expectation derived from 
the scaled nucleon-nucleon and nucleon-nucleus reactions.

These experimental results are
consistent with particle production occurring from 
a strangeness dense source in which the strange quarks 
are already  made, are freely moving around and are 
readily available. This is by definition, the
deconfined state we refer to as the quark-gluon plasma. 
We are not aware of a consistent interpretation of the rich and diverse
experimental results  not using QGP formation.
It is important here to remember that since 
there are many results on strange hadron 
production, any alternative description 
 has to be tested on all data available.

We think that strangeness enhancement has turned out to be
a very practical signature of quark-gluon plasma physics.  
This report summarizes our work of past 18 months in
this area -- for further details the reader may consult
our recent articles.~\cite{Let00,Ham00,Raf00,Tor00}
In the following section \ref{QGP} 
we will describe our most recent advances
in the study of properties of QGP at time of its formation and its
sudden breakup.~\cite{Ham00} The initial  conditions are determining the
overall strangeness yield. We use the properties of QGP at sudden
breakup in the following section \ref{analyze}, testing the 
consistency of the chemical freeze-out condition with the 
properties of the QGP.~\cite{Let00}   Recently,~\cite{Raf00} 
we have been able to  understand 
the mechanism of sudden QGP hadronization 
in terms of  a mechanical instability arising within a 
supercooled and rapidly  expanding fireball,
 as we will discuss in section \ref{instability}.

An important experimental observation
pointing to sudden  QGP breakup is that 
strange baryons and antibaryons  have very similar
$m_\bot$-spectra~\cite{Ant00} and are thus produced by the 
same mechanism.~\cite{RD87}  Moreover, as we will show in
section \ref{thermal}, these spectra imply that the
  thermal freeze-out  of these
rarely produced particles  occurs together with the chemical
(abundance) freeze-out.~\cite{Tor00} This occurs for all collision
centralities measured by experiment WA97.

In recent months first strangeness results have been obtained 
at RHIC. Our analysis effort regarding these results has just 
begun. Some of our RHIC predictions~\cite{Raf99,Bas00}
can already be compared with experiment, and we will show 
in section~\ref{RHIC} to be in agreement with our picture  of how 
strangeness and QGP evolves at RHIC.

\section{QGP fireball} \label{QGP}
\subsection{Formulation of the QGP model}
Using the latest lattice-QCD results, we have been able to better 
model the properties of the QGP,~\cite{Ham00} compared to our earlier 
considerations.~\cite{Acta96}
The key ingredients of our current approach are:
\begin{enumerate}
\item
We relate the QCD scale to the temperature $T=1/\beta$, 
 we use for the scale the Matsubara frequency~\cite{Pes00}
($\mu$  without a subscript is {\it not} a chemical potential, it is
the QCD scale):
\begin{equation}
\label{runalTmu}
\mu=2\pi \beta^{-1}\sqrt{1+\frac{1}{\pi^2}\ln^2\lambda_{\mathrm q}}
=2\sqrt{(\pi T)^2+\mu_{\mathrm q}^2}\,.
\end{equation}
This extension to finite chemical  potential $\mu_{\mathrm q}$, or 
equivalently quark fugacity  $\lambda_{\mathrm q}=\exp{\mu_{\mathrm q}/T}$, 
is motivated by the form of plasma frequency  entering
the computation of the vacuum polarization function.~\cite{Vij95}
\item
We obtain the interacting strength $\alpha_s(\mu)$
integrating numerically the renormalization group equations 
\begin{equation}
\label{dalfa2loop}
\mu \frac{\partial \alpha_s}{\partial \mu}=
-b_0\alpha_s^2-b_1\alpha_s^3+\ldots \equiv \beta^{\mbox{\scriptsize pert}}_2\,.
\end{equation}
$\beta^{\mbox{\scriptsize pert}}_2$ is
the beta-function of the renormalization group 
in two loop approximation, and 
$$b_0=\frac{11-2n_{\mathrm f}/3}{2\pi}\,,\quad 
   b_1=\frac{51-19n_{\mathrm f}/3}{4\pi^2}\,.$$ 
$\beta^{\mbox{\scriptsize pert}}_2$
does not depend on the renormalization scheme,
and solutions of Eq.\,(\ref{dalfa2loop}) differ from higher 
order renormalization scheme
dependent results  by less than the error introduced by the experimental 
uncertainty in the measured value of $\alpha_s(\mu=M_Z)=0.118+0.001-0.0016$. 
\item
We introduce, in the domain of freely mobile quarks 
and gluons, a finite vacuum energy  density:
$${\cal B}=0.19\,\frac{\mbox{GeV}}{\mbox{fm}^3}\,.$$
This also implies, by virtue of relativistic invariance,
that there must be a (negative) 
associated pressure acting on the surface of this volume, 
aiming to reduce the size of the deconfined region.  
These two properties of the vacuum follow
consistently from the vacuum partition function:
\begin{equation}
\label{Zbag}
\ln{\cal Z}_{\mbox{\scriptsize vac}}\equiv -{\cal B}V\beta\,.
\end{equation}
\item 
The partition function of the quark-gluon liquid comprises interacting 
gluons, $n_{\mathrm q}$ flavors of light quarks,~\cite{Chi78} 
and the vacuum ${\cal B}$-term. We
incorporate further the strange quarks by assuming that their mass 
in effect reduces their effective number  $n_{\mathrm s}<1$:
\begin{eqnarray}
\label{ZQGPL}
\frac{T}{V}\ln{\cal Z}_{\mathrm QGP}
\equiv P_{\mathrm QGP}
&&=
-{\cal B}+\frac{8}{45\pi^2}c_1(\pi T)^4  \nonumber\\ 
&&+
\frac{n_{\mathrm q}}{15\pi^2}
\left[\frac{7}{4}c_2(\pi T)^4\!+\frac{15}{2}c_3\!\left(\!
\mu_{\mathrm q}^2(\pi T)^2 + \frac{1}{2}\mu_{\mathrm q}^4
\right)\right]
\nonumber\\ 
&&+
\frac{n_{\mathrm s}}{15\pi^2}
\left[\frac{7}{4}c_2(\pi T)^4\!+\frac{15}{2}c_3\!\left(\!
\mu_{\mathrm s}^2(\pi T)^2 + \frac{1}{2}\mu_{\mathrm s}^4
\right)\right]\,,
\end{eqnarray}
 where:
\begin{eqnarray}
\label{ICZQGP}
c_1=1-\frac{15\alpha_s}{4\pi}+ \cdots,\quad 
c_2=1-\frac{50\alpha_s}{21\pi}+ \cdots,\quad
c_3=1-\frac{2\alpha_s}{\pi}+ \cdots.
\end{eqnarray}
\end{enumerate}

\subsection{Properties of QGP-liquid}
\begin{figure}[bt]
\vspace*{-0.2cm}
\centerline{
\hspace*{3.5cm}\epsfig{width=12cm,clip=,figure=\pathnow 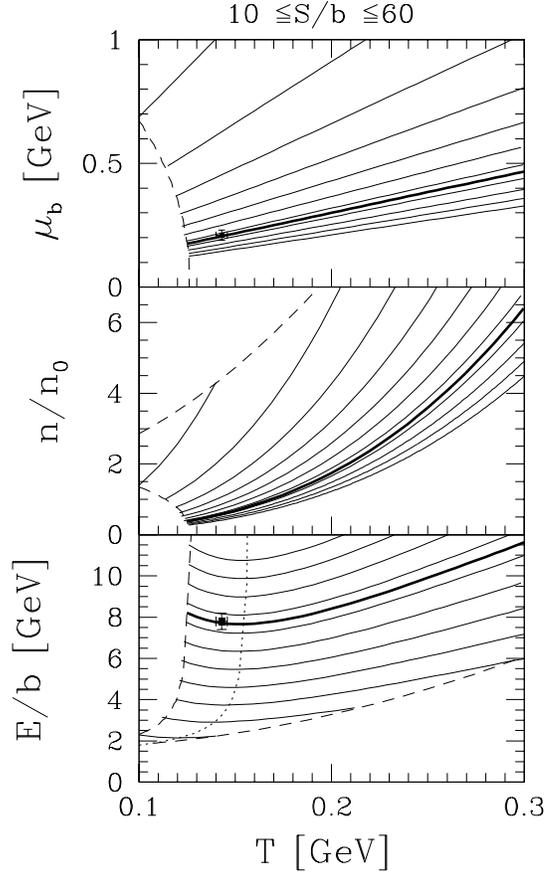}
}
\vspace*{-0.6cm}
\caption{ 
 From top to bottom: $\mu_b,\ n/n_0$ and $E/B$; 
lines shown correspond to fixed entropy per baryon 
$S/B= 10$ to 60 by step of 5 (left to right). 
Thick solid lines: result for $S/B=42.5$.
Limits:  energy density $\varepsilon_{\mathrm q,g}=0.5$\,GeV/fm$^3$ and
baryo-chemical potential $\mu_b=1$\,GeV. The experimental points denote
chemical freeze-out analysis result.~\protect\cite{Let00}
\label{EOSSfix}
}
\end{figure}

We show properties of the quark-gluon liquid defined by Eq.\,\ref{Zbag} 
in a wider range of parameters
at fixed entropy per baryon $S/B$, in the range $S/B=10$--60 
 in step of 5 units. In the  top panel  in 
figure \ref{EOSSfix}, we show  
baryo-chemical potential $\mu_b$, in middle panel
baryon density $n/n_0$, here  $n_0=0.16/\mbox{fm}^3$, and bottom left
the energy per baryon $E/B$. In top and middle 
 panel the low entropy results are top-left 
in figure, in bottom panel bottom left. 
The  highlighted curve, in figure \ref{EOSSfix},
is for the  value $S/B=42.5$\,, applicable to the 158 $A$GeV Pb--Pb 
collisions. The dotted line, at the 
minimum of $E/B\vert_{S/B}$, is where the vacuum 
and quark-gluon gas pressure balance. 

Little entropy is produced during the nearly adiabatic  evolution of 
the QGP fireball.~\cite{entro} Thus
the lines in the lower panel of figure \ref{EOSSfix}
characterize the approximate trajectory in time of the fireball. After
initial drop in energy per baryon due to transfer of energy to accelerating 
expansion of the fireball, during the deep supercooling process, the motion 
is slowed and thus energy per baryon increases. 
The thick line is our expectation for the fireball 
made in Pb--Pb interactions at the projectile energy 158$A$ GeV. 
The cross shows the result of  chemical freeze-out 
analysis presented in next section \ref{analyze}. \cite{Let00}

\begin{figure}[tb]
\vspace*{-1.7cm}
\centerline{\hskip 0.5cm
\hspace*{1.0cm}\epsfig{width=10.cm,clip=,figure=\pathnow 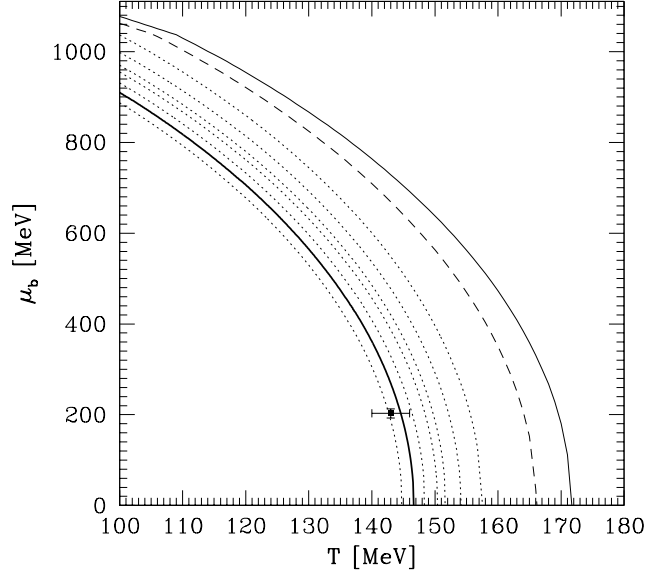}
}
\vspace*{-0.8cm}
\caption{ 
Thin solid and dashed lines: equilibrium phase transition from
hadron gas to QGP liquid without and with excluded volume correction,
respectively.  Dotted:  
breakup condition at shape parameter $\kappa=0.6$, for expansion velocity 
$v_{\mbox{\scriptsize c}}^2=0, 1/10, 1/6, 1/5, 1/4$ and 1/3, and thick line for 
$v_{\mbox{\scriptsize c}}=0.54$. The experimental point denotes 
chemical non-equilibrium freeze-out analysis result.~\protect\cite{Let00}
\label{PLMUPLIQ}
}
\end{figure}

We also have compared the chemical freeze-out conditions with the
phase  transition properties. 
The hadron gas behavior is obtained evaluating and summing 
the contributions  of  all known hadronic resonances considered to be
point particles. 
The thin solid line in the $T,\mu_{\mbox{\scriptsize b}}$ plane  
in figure \ref{PLMUPLIQ} shows  where the pressure of the quark-gluon
liquid  equals the equilibrated hadron gas pressure. 
When we allow for 
finite volume of hadrons, \cite{HR80} we find that the hadron  pressure is
slightly reduced, leading to some (5\,MeV) reduction in the  equilibrium transition 
temperature, as is shown by  the dashed line in figure~\ref{PLMUPLIQ}\,.
For vanishing baryo-chemical potential, 
we note in figure \ref{PLMUPLIQ} that the equilibrium
phase transition temperature is  $T_{\mbox{\scriptsize pt}}\simeq 172$\,MeV,
and when finite hadron size is allowed, $T_{\mbox{\scriptsize fp}}\simeq 166$\,MeV,
The scale in temperature we discuss is result of 
comparison with lattice gauge results. 
Within the lattice calculations~\cite{Kar00}, 
it arises from the comparison with  the string tension.

The dotted lines, in figure \ref{PLMUPLIQ}, show where the flowing QGP liquid 
has a pressure which balances the vacuum pressure, and thus these lines
correspond to sudden break up at a velocity  for (from right to left) 
$v_{\mbox{\scriptsize c}}^2=0, 1/10, 1/6, 1/5, 1/4$ and $1/3$, 
see section~\ref{instability}, specifically the dotted lines 
in figure \ref{PLMUPLIQ} correspond to 
the condition Eq.\,(\ref{BPeqv}) using the shape parameter 
$\kappa=0.6$, Eq.\,(\ref{Pn}).  The
last dotted line corresponds thus to an expansion flow with 
the velocity of sound of  relativistic noninteracting massless gas. The thick
solid line corresponds to an expansion with $v_{\mbox{\scriptsize c}}=0.54$.  
The hadron analysis result is also shown as 
presented in section~\ref{analyze}.~\cite{Let00} 
Comparing in figure \ref{PLMUPLIQ} 
thin solid/dashed with the  thick line,
we recognize the deep supercooling as required for the explosive 
fireball disintegration. The super-cooled  zero pressure $ P=0$ 
baryonfree QGP   temperature is at $T_{\mbox{\scriptsize sc}}=157$\,MeV, 
(see the intercept of the first dashed line to the right 
in figure \ref{PLMUPLIQ}) and an expanding fireball can deeply 
super-cool to $T_{\mbox{\scriptsize dsc}}\simeq 147$\,MeV 
(see the intercept of thick solid line)
before the mechanical instability occurs.

\subsection{Initial conditions in QGP formation}

\begin{figure}[tb]
\vspace*{-3.6cm}
\centerline{\hspace*{0.20cm}\psfig{width=12.5cm,clip=,figure=\pathnow 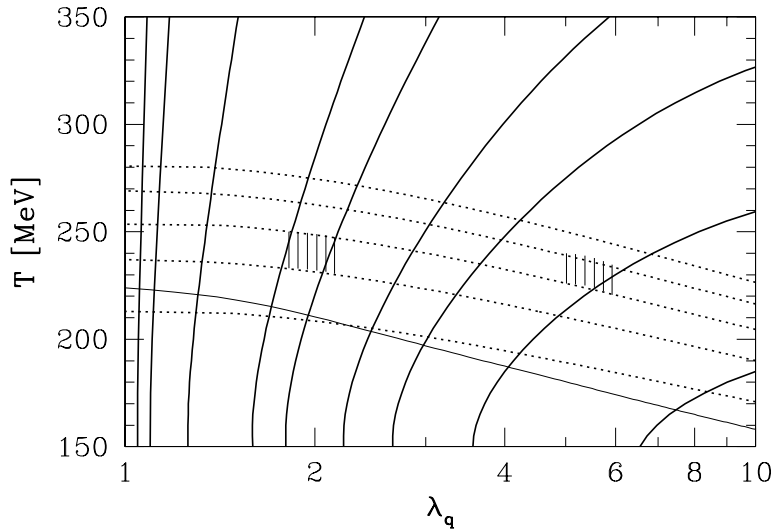}}
\vspace*{-0.6cm}
\caption{ 
Contours of energy per baryon in QGP in the $T$--$\lambda_q$ plane 
for $n_f=1$: From right to left 
$E/B$=3, 4, 5, 6, 8, 10, 20, 50 and 100\,GeV\,.
Thin, nearly horizontal line: hadronic gas phase has the same pressure 
as the QGP-liquid with semi-equilibrated quark flavor. Dotted lines from
bottom to top: pressure in QGP liquid equals 20\%,\,40\%,\,60\%,\,80\%,\, 
and 100\%, of the dynamical collisional pressure.
\label{Ebnf1fix}
}
\end{figure}

The formation of the QGP occurs well before chemical
equilibration of quarks. We model this situation varying $n_f$.
For $n_f=1$ we show, in figure \ref{Ebnf1fix}, lines
of fixed energy per baryon $E/B=$3,\,4,\,5,\,6,\,8,\,10,\,20,\,50 and 100\,GeV, The 
horizontal solid line is where the equilibrated hadronic gas phase has the same 
pressure as QGP-liquid with  semi-equilibrated  quark abundance. The 
free energy of the QGP liquid must be lower (pressure higher) 
in order for  hadrons to dissolve into the plasma phase. The dotted lines  
in figure \ref{Ebnf1fix}, from bottom to 
top, show where the pressure of the semi-equilibrated QGP phase is equal to 
$\eta=$ 20\%,\,40\%,\,60\%,\,80\% and 100\%,\, $\eta$
being the `stopping' fraction of the dynamical collisional pressure \cite{Acta96}:
\begin{equation}
 \label{stop}
P_{\mathrm{col}}=\eta\rho_0\frac{P_{\mathrm{CM}}^2}{E_{\mathrm{CM}}}\,.
\end{equation}

The rationale to study, in figure \ref{Ebnf1fix}, 
lines at fixed $E/B$ is that, 
during the nuclear collision which lasts about 
$2R_N/\gamma_L2c\simeq 13/18$\,fm/$c$, where 
$\gamma_L$ is the Lorentz factor between 
the lab and CM frame and $R_N$ is the nuclear 
radius, parton collisions 
lead to a partial (assumed here to be 1/2) 
chemical equilibration of the hadron 
matter. At that time,  the pressure exercised 
corresponds to collisional pressure
$P_{\mathrm{col}}$\,. This stopping fraction, 
seen in the transverse energy produced, 
is about 40\% for S--S collisions 
at 200$A$ GeV and 60\% for Pb--Pb collisions 
at 158$A$ GeV. If the momentum-energy
and baryon number stopping are similar, 
as we see in the experimental 
data, then the SPS collisions at 
160--200$A$\,GeV are found in the highlighted
area left of center of the figure. In the middle 
of upper boundary of this area, we
would expect the beginning evolution of the thermal but not yet 
chemically equilibrated Pb--Pb fireball, and in the lower
left corner of the S--S fireball. We note that 
the temperature reached in S--S 
case is seen to be about 25 MeV lower than in the Pb--Pb case.
The lowest dotted line (20\% stopping) nearly coincides
 with the non-equilibrium phase
boundary (solid horizontal line, in figure \ref{Ebnf1fix}) 
and thus we conclude  that this is, for the condition 
$n_f=1$, the lowest
stopping that can lead to formation of a deconfined 
QGP phase. Such a low stopping would
be encountered possibly in lighter than S--S 
collision systems or/and at large impact 
parameter interactions of larger nuclei.

The highlighted area, right
of center of the figure \ref{Ebnf1fix}, 
corresponds to the expected conditions in Pb--Pb 
collisions at 40$A$ GeV. If we assume that the stopping here is near 80\%,
then the initial conditions for fireball evolution would be found towards the 
upper right corner of this highlighted area. We recognize that the
higher stopping nearly completely compensates the effect of reduced 
available energy in the collision and indeed, we expect that we form QGP 
also at these collision energies. It is important to realize that
we are entering a domain of parameters, in particular $\lambda_q$, 
for which the extrapolation of the  lattice results is not 
necessarily reliable, and thus our equations of state have
increased  systematic uncertainty.

\section{Fermi-model data analysis}\label{analyze}
\subsection{Properties of hadronic matter fireball}
A full account of our prior analysis of the 158$A$ GeV Pb--Pb collision
system has appeared.~\cite{Let00} We briefly summarize the results
that we require for the study of QGP properties at fireball breakup. 
In table \ref{fitqpbs}, in upper section, we present the parameters
$T_f$ (the chemical freeze-out temperature), $v_c$ (the collective flow 
velocity at sudden breakup), $\lambda_{q}$ (the light quark fugacity),
$\lambda_{s}$  (the strange quark fugacity), $\gamma_{q}$ (the light
quark phase space occupancy), $\gamma_{s}$ (the strange quark
phase space occupancy). These chemical properties 
are derived from analysis of all hadrons 
excluding $\Omega$ and $\overline\Omega$, which data 
points are not following the same systematic production pattern. 
These parameters characterize completely the physical properties of the 
produced hadrons, and these properties
are shown in the bottom section of table \ref{fitqpbs}.

In the heading of the table, the
total error, $\chi^2$ is shown, along with the number of data points $N$, parameters 
$p$ and data point constraints $r$. The confidence level that is
reached in our description is near or above 90\%, depending on 
scenario considered.  
The scenarios we consider are seen in the
columns of table \ref{fitqpbs}: an unconstrained
description of all data in the first column, constraint to exact strangeness
conservation in the observed hadrons, second column. Since in both
cases the parameter $\gamma_{q}$ assumes value that maximizes the 
entropy and energy content in the pion gas, we assume this value 
in the so constrained third column.

We can now  check the consistency between the statistical parameters 
(top panel of table~\ref{fitqpbs}) and the 
physical properties of the fireball (bottom  panel of table \ref{fitqpbs})
which are maintained in the process of 
hadronization.  We note that the energy shown in this table, is the intrinsic 
energy in the flowing frame. The CM-laboratory energy includes the kinetic 
energy of the flow and thus is greater, to be  obtained 
multiplying the result shown in table \ref{fitqpbs} 
by the Lorenz factor $\gamma=1/\sqrt{1-v^2_c}=1.19$.
Thus the initial value of the energy per baryon that the 
system had before expansion started has been $E^0/B\simeq 9.3$\,GeV.
\begin{table}[t]
\caption{\label{fitqpbs}
Results of study of Pb--Pb hadron production:~\protect\cite{Let00} 
in the heading: the total quadratic relative error 
$\chi^2_{\rm T}$, number of data points $N$, parameters $p$ and 
redundancies $r$;  in the upper section: statistical model parameters
which best describe the experimental results for Pb--Pb data.
Bottom section: specific energy, entropy, anti-strangeness, net strangeness
 of  the full hadron phase space characterized by these
statistical parameters. In column one, all statistical parameters and 
the flow vary. In column two, we fix $\lambda_{s}$ by requirement of 
strangeness conservation, and in column three, we fix $\gamma_{q}$ at
the pion condensation point $\gamma_{q}=\gamma_{q}^c$.}
\begin{center}
\begin{tabular}{l|ccc}
                       & Pb$|_v$            & Pb$|_v^{\rm sb}$ & Pb$|_v^{\rm sc}$     \\
$\chi^2_{\rm T};\ N;p;r$&2.5;\ 12;\,6;\,2   & 3.2;\ 12;\,5;\,2 & 2.6;\ 12;\,5;\,2     \\
\hline
$T_{f}$ [MeV]          &    142 $\pm$ 3     &  144 $\pm$ 2     &  142 $\pm$ 2       \\
$v_c$                  &   0.54 $\pm$ 0.04  & 0.54 $\pm$ 0.025 & 0.54 $\pm$ 0.025   \\
$\lambda_{q}$          &   1.61 $\pm$ 0.02  & 1.605 $\pm$ 0.025& 1.615 $\pm$ 0.025   \\
$\lambda_{s}$          &   1.09 $\pm$ 0.02  & 1.10$^*$         & 1.09 $\pm$ 0.02      \\
$\gamma_{q}$           &   1.7 $\pm$ 0.5    & 1.8$\pm$ 0.2   &${\gamma_{q}^c}^*=e^{m_\pi/2T_f}$ \\
$\gamma_{s}/\gamma_{q}$&   0.79 $\pm$ 0.05  & 0.80 $\pm$ 0.05  & 0.79 $\pm$ 0.05     \\
\hline
$E_{f}/B$              &   7.8 $\pm$ 0.5    & 7.7 $\pm$ 0.5    & 7.8 $\pm$ 0.5     \\
$S_{f}/B$              &    42 $\pm$ 3      & 41 $\pm$ 3       & 43 $\pm$ 3        \\
${s}_{f}/B$            &  0.69 $\pm$ 0.04   & 0.67 $\pm$ 0.05  & 0.70 $\pm$ 0.05    \\
$({\bar s}_f-s_f)/B\ \ $   &  0.03 $\pm$ 0.04   & 0$^*$            &  0.04 $\pm$ 0.05   \\
\end{tabular}
\end{center}
\end{table}

\subsection{Interpretation in terms of QGP}
In the bottom panel in figure~\ref{EOSSfix}, we saw that the Temperature 
$T_f=143\pm3$\,MeV and intrinsic (frame at rest)
 energy per baryon $E/B=7.8$\,GeV 
where just at $S/B=42.5$ seen  table \ref{fitqpbs}. Similarly, in the top 
panel, the baryo-chemical potential $\mu_b=3 T_f\ln \lambda_q=204\pm10$\,MeV 
is  as required for the consistency of QGP properties. 
The fireball hadron freeze-out and/or breakup condition
is thus found   well below  
the QGP to HG phase transition temperature. 
Both the specific energy and entropy
content of the fireball are consistent with the statistical parameters 
$T_f$ and $\mu_b$ according to our equations of state of the quark-gluon 
liquid. We can also evaluate the hadronic phase space
energy density. We allow the  excluded volume 
correction.~\cite{HR80} Considering that the point particle 
phase space energy density $\varepsilon_{pt}= 1.1$\,GeV/fm$^3$, we obtain
$\varepsilon_{HG}\simeq 0.4$\,GeV/fm$^3$ at chemical freeze-out, 
using the  value of  ${\cal B}=0.19$\,GeV/fm$^3$. This energy density is 
neraly the same as one finds in QGP. Thus we see that the
chemical freeze-out is due to the direct fireball breakup.

\begin{figure}[tb]
\vspace*{1.9cm}
\centerline{\hspace*{-.5cm}
\psfig{width=10cm,figure=\pathnow 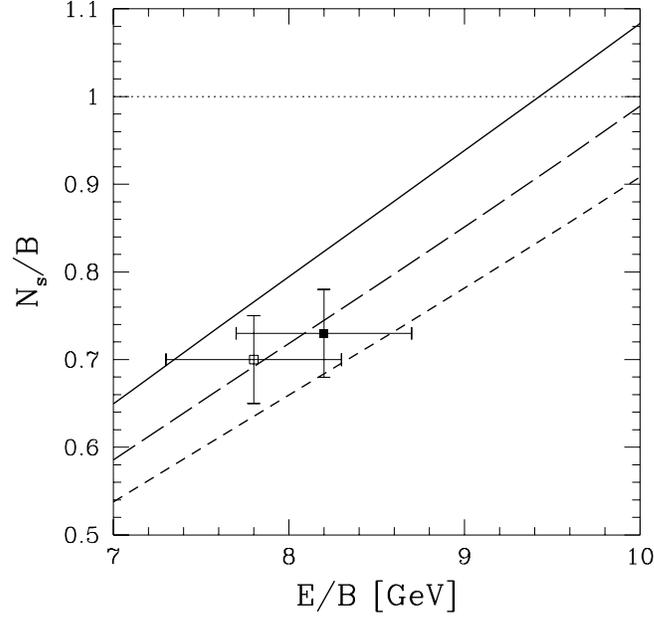}
}
\vspace*{0.2cm}
\caption{ 
QGP fireball specific per baryon strangeness abundance  
as function of $E/B$ energy per baryon
in the fireball, for running  
$\alpha_{\rm s}(M_Z)=0.118$ and $m_{\rm s}(1GeV)=200$\,MeV,
corresponding to $m_{\rm s}(M_Z)=90$\,MeV.
Solid lines: for Pb--Pb, stopping 57\% and 360 participants, 
long-dashed lines: for Ag--Ag, stopping 54\% and 180 participants,
short-dashed lines: for S--Au/W/Pb stopping 52\% 
and 90 participants in QGP fireball. The solid square is the
result of an analysis for S-Au/W/Pb 200$A$ GeV reaction system, 
and open square for Pb--Pb at 158$A$ GeV, as shown in the
lower section of table~\ref{fitqpbs}. 
\label{figsbrunE}}
\end{figure}

More generally, the chemical analysis evidence 
for the QGP nature of the fireball
is as follows:~\cite{Let00}\\
{\bf  a)} The value of strange quark 
fugacity $\lambda_{s}$ we find in our analysis can also  be obtained  
from the  requirement that strangeness  balances,
$\langle N_{s}-N_{\bar s}\rangle=0\,,$ in QGP,
which for a source in which all $s$ an $\bar s$ quarks are unbound and thus 
have symmetric phase space, implies $\lambda_{s}=1$\,.
 However, the Coulomb distortion
of the strange quark phase space plays an important role in the
understanding of this constraint for Pb--Pb collisions,~cite{Raf91}
leading to the Coulomb-deformed value $\lambda_{s}\simeq 1.1$, which is identical 
to the value obtained from experimental data analysis $\lambda_s^a=1.09\pm0.02$\,.\\
{\bf  b)} The  phase space occupancy of light quarks 
$\gamma_{q}$ even if it were before gluon fragmentation near or at 
the equilibrium value $\gamma_{q}=1$, must increase above this
equilibrium value.  There is an upper limit:
$\gamma_{q}<\gamma_{q}^c\equiv e^{m_\pi/2T_f}\simeq 1.67$\,,
which arises to maximize the entropy density in the confined hadron phase. In
our analysis we find values in this narrow interval $1\le \gamma_{q}\le \gamma_{q}^c$ \\
{\bf  c)} The strange quark phase space occupancy 
$\gamma_{s}$ we compute within the framework 
of kinetic theory where  strangeness pair production arises 
in gluon fusion,~\cite{Raf82}. The result depends on temperature reached 
 in early stages of the collision, and following
dilution effect in which the already produced strangeness can even over saturate the
`thiner' low temperature phase space. Moreover, some gluon fragmentation also
enriches $\gamma_s$ as measured by hadron abundance, and thus we find $gmma_s\ge 1$. 
We note that  the ratio $\gamma_s/\gamma_q$ has been earlier confounded 
with $\gamma_s$, which therefore
is stated in  table \ref{fitqpbs}.\\
{\bf  d)} The 
strangeness yield,  $N_s/B=N_{\bar s}/B\simeq 0.7$, predicted early on as the 
result of QGP formation~\cite{Raf82} is also one of the
results of  data analysis seen in table \ref{fitqpbs}. This result we 
obtain within the modern kinetic model of strangeness
production in QGP.~\cite{Let00} The results are shown in 
figure~\ref{figsbrunE} as function of the intrinsic specific energy available 
in the fireball $E/B$, for the three collision systems S-Au/W/Pb (short-dashed
line), Ag--Ag (long dashed) and Pb--Pb (solid line). 
The solid square is the `experimental' result obtained in our analysis of the 
 S-Au/W/Pb system, and the open square is for Pb--Pb. The experimental
value of stopping we employ (see legend to figure \ref{figsbrunE})
were obtained by NA49 collaboration,~\cite{NA49stop}.

Stopping enters the computation of 
the initial temperature from the collision energy, see Eq.\,(\ref{stop}). 
We extrapolate as function of $E/B$, assuming that the 
stopping fraction for the collisional pressure is known, and as here
assumed, not changing with collision energy in the narrow
interval shown in figure~\ref{figsbrunE}, leading to a steep
rise of strangeness production with collision energy.

\section{Supercooling and sudden breakup}\label{instability}
\subsection{Instability condition}
Sudden hadronization of dense matter fireballs formed in  
158$A$ GeV  Pb--Pb collisions requires 
that a  global  instability of the dense hadronic matter fireball
is encountered in its evolution. 
Given the hadronization temperature we found, $T\simeq 143$\,MeV,
which is about 15\,MeV below the equilibrium phase transition 
of QGP phase to hadron gas it is
a natural step to consider the possibility of supercooling of the QGP
phase. As many of us have experienced in diverse daily life experiences,
once the supercooled phase is disturbed, 
a truly sudden break-up occurs. QGP instability  was 
suggested by  Cs\"org\H{o} and Csernai~\cite{Cso94} to explain
sudden hadronization.~\cite{Raf91}
Despite extensive ensuing study of 
potential microscopic mechanisms,~\cite{Cse95,Bir99,Mis99} 
a convincing elementary instability  mechanism has not been found.

As the fireball expands, energy is transfered from thermal motion
to collective flow in adiabatic fashion. At a certain point 
in deep supercooling, energy flow is reversed, and the matter
flow begins to slow down. However, if the flow velocity remains 
large but the surface pressure ${\cal P}\to 0$, continued expansion
must rip the fireball apart. This is the natural mechanical 
instability we proposed as the sudden hadronization 
condition.~\cite{Raf00} This simple mechanism does not require 
a microscopic process for the onset of instability.

We consider the exploding fireball dynamics 
in its center of momentum frame of reference.
The surface normal vector of exploding fireball 
is $\vec n$, and the local velocity 
of matter flow $\vec v_{\mbox{\scriptsize c}}$. 
The rate of  momentum flow vector $\vec {\cal P}$ 
at the surface  is obtained from the energy-stress tensor $T_{kl}$:~\cite{LanHyd} 
\begin{equation}\label{Peqv}
\vec {\cal P}=P^{(i)}\vec n+(P^{(i)}+\varepsilon^{(i)})
  \frac{\vec v_{\mbox{\scriptsize c}}\, \vec v_{\mbox{\scriptsize c}}\!\cdot\! \vec n}
          {1-\vec v_{\mbox{\scriptsize c}}^{\,2}}\,.
\end{equation}
The upper index  $(i)$ refers for the intrinsic energy 
density $\varepsilon$ and
pressure $P$  of matter in the  frame of reference,  
locally at rest, {\it i.e.,} observed by a co-moving observer.  
We  omit the superscript  $(i)$ in the following.
 For the fireball expansion to continue, ${\cal P}\equiv |\vec {\cal P} |> 0$
is required.  For ${\cal P}\to 0$ at $v_{\mbox{\scriptsize c}}\ne 0$, we have
a conflict between the desire of the motion to stop or even reverse, and 
the continued inertial expansion. 

When the flow velocity remains 
large but ${\cal P}\to 0$, the intrinsic 
pressure $P$ must be negative. As 
illustration consider the fireball to be made of 
a quark-gluon liquid confined by an external vacuum 
pressure $\cal B$.  The total pressure 
and energy  comprise particle (subscript $\mbox{\scriptsize p}$)
and the  vacuum properties:
\begin{equation}\label{EPB1}
P=P_{\mbox{\scriptsize p}}-{\cal B}\,,\quad \varepsilon
 =\varepsilon_{\mbox{\scriptsize p}}+{\cal B}\,.
\end{equation}
Eq.\,(\ref{Peqv}) with $\vec {\cal P}=0$ thus reads:
\begin{equation}\label{BPeqv}
{\cal B}\vec n=P_{\mbox{\scriptsize p}}\vec n+
      (P_{\mbox{\scriptsize p}}+\varepsilon_{\mbox{\scriptsize p}})
\frac{\vec v_{\mbox{\scriptsize c}} \, \vec v_{\mbox{\scriptsize c}}\!\cdot \!\vec n}
        {1-v_{\mbox{\scriptsize c}}^{2}}\,,
\end{equation}
and it describes the (equilibrium) condition where the pressure of the expanding 
quark-gluon  fluid is just balanced by the external vacuum pressure.

Expansion beyond  ${\cal P}\to 0$ is in general not possible. 
A surface region of the fireball that  reached it but 
continues to flow outwards 
must   be torn apart. This is a collective instability and thus the 
ensuing disintegration of the fireball matter will be 
very rapid, provided that much of the surface reaches this condition.
We adopt the condition $\vec {\cal P}=0$ at any surface region to be
the  instability condition of an expanding hadron 
matter fireball.

Negative internal pressure $P<0$ is a requirement. 
At this stage the fireball must thus
be significantly supercooled.
The adiabatic transfer of internal heat  into accelerating 
flow of matter provides the mechanism which 
leads on the scale of $\tau=2\,10^{-23}$\,s to the development of this `deep'
supercooling.

\subsection{Consistency with QGP properties}
The sudden break up due to mechanical instability of course
will occur for all phases of elementary matter and thus we
consider next in how far the observed freeze-out conditions and 
other features of hadronization process are consistent with the 
QGP mechanical breakup, and explore if experiments show evidence for
supercooling. 

It is possible to determine experimentally if
the  condition $P<0$ has been reached. Namely, the
 Gibbs-Duham relation for a unit volume:
\begin{equation}\label{1stlaw}
P=T\sigma+\mu_{\mbox{\scriptsize b}}\nu_{\mbox{\scriptsize b}}- \varepsilon\,,
\end{equation}
relates   the pressure, to  entropy 
density $\sigma=S/V$\,, energy density $\varepsilon=E/V$\,, 
and baryon density $\nu_{\mbox{\scriptsize b}}=b/V$\,, $V$ is the volume,
$T$ is the temperature, and $\mu_{\mbox{\scriptsize b}}$ the baryochemical potential.
Dividing by $\varepsilon$ we obtain:
\begin{equation}\label{EBStest}
\frac{PV}{E}=\frac{T_{\mbox{\scriptsize h}}}{E/S}
    +\frac{\mu_{\mbox{\scriptsize b}}}{E/B}-1\,.
\end{equation}
The microscopic processes governing the fireball
breakup determine how the quantities entering the right hand side of 
Eq.\,(\ref{EBStest}) are changed as hadrons emerge. Understanding this
we can determine, if the intrinsic fireball pressure  prior to  breakup,  
has been negative. 

The energy $E$  and baryon content  $b$ of the fireball are  conserved. 
Entropy $S$ is conserved when the gluon content of a QGP 
fireball is transformed into 
quark pairs in the entropy conserving  process 
$G+G\to q+\bar q$. Similarly, when
quarks and antiquarks recombine into hadrons,  entropy is conserved in the 
range of parameters of interest here.  Thus  also $E/B$ and $S/B$ is conserved 
across hadronization condition. The sudden hadronization process 
also maintains the temperature $T$ and baryo-chemical potential $\mu_b$
across the phase boundary. What changes in QGP breakup 
are the chemical occupancy parameters. 
As gluons convert into quark pairs and hadrons $\gamma_g\to 0$ but
the number occupancy of light valance quark  pairs 
increases $\gamma_q>\gamma_{q_0}\simeq 1$
 increases significantly, along with  the number occupancy 
of strange quark  pairs $\gamma_s>\gamma_{s_0}\simeq 1$.

Evaluating Eq.\,(\ref{EBStest}) using the results of the data analysis
presented above, 
we indeed obtain $P_f<0$. The magnitude of $|P_f|$  can vary between 
a few percent (in terms of energy density $E_f/V$), up to 20\% for the latest 
published result.~\cite{Let00} The
precise value, which arises from several cancellations of larger numbers 
is sensitive to the strategy of how the currently available experimental 
data is described, {\it e.g.,} if strangeness  conservation is 
implemented, and  if so, if  differentially at
each rapidity, or as an overall conservation law; 
how many high mass resonances 
can be excited in hadronization process, etc ...

Understanding in detail the breakup  condition ${\cal P}\to 0$ requires
that we model the shape and direction of flow in the late
stage of fireball evolution, obviously not an easy task. However,
considering $\vec n \!\cdot\! \vec {\cal P}\to 0$, we  find  
the constraint:
\begin{equation}\label{Pn}
\frac{-PV/E}{1+PV/E}=\kappa \frac{v_{\mbox{\scriptsize c}}^2}
               {1-v_{\mbox{\scriptsize c}}^2}\,, \quad 
  \kappa=(\vec v_{\mbox{\scriptsize c}}\! \cdot\! \vec n)^2/v_{\mbox{\scriptsize c}}^2\,.
\end{equation}
For an exactly symmetrical, spherical expansion the two vectors
$\vec v_{\mbox{\scriptsize c}}$ and $\vec n$  are everywhere parallel, thus $\kappa\to 1$.
However, in 158$A$ GeV Pb--Pb reactions  the longitudinal flow is 
considerably greater than the transverse flow,~\cite{NA49stop} and we note
$\kappa\to 0$ for a longitudinally evolving cylindrical 
fireball. For the Pb--Pb collisions considered here, our analysis 
suggest $0.1<\kappa<0.6$.

We now substitute, in Eq.\,(\ref{Pn}), the fireball matter properties 
employing the  Gibbs-Duham relation, Eq.\,(\ref{1stlaw}), and arrive at: 
\begin{equation}\label{EBSfinal}
\frac{E}{S}=\left(T_{\mbox{\scriptsize h}}+\frac{\mu_{\mbox{\scriptsize b}}}{S/B}\right)
     \left\{1+\kappa \frac{v_{\mbox{\scriptsize c}}^2}
                 {1-v_{\mbox{\scriptsize c}}^2}\right\}\,.
\end{equation}
Eq.\,(\ref{EBSfinal}) establishes a general 
constraint characterizing the fireball breakup condition.

Deep supercooling requires a  first order phase transition, and
this in turn implies presence of latent heat $\cal B$. 
Physical consistency then requires  presence of 
external (negative) vacuum pressure $-{\cal B}$. 
We now combine the theoretical properties of the QGP equations of state 
with the dynamical fireball properties in order to constrain ${\cal B}$. 
Reviewing Eq.\,(\ref{EBStest}), we obtain:
\begin{equation}\label{EBSres}\label{EBSineq}
-\frac{PV}{E}\varepsilon_{\mbox{\scriptsize QGP}}
+P_{\mbox{\scriptsize p}}={\cal B}\,,
\end{equation}
To evaluate $\cal B$, we note that lattice results for $\varepsilon_{\mbox{\scriptsize QGP}}$ 
are well represented by $\varepsilon_{\mbox{\scriptsize QGP}}=aT^4$,  with $a\simeq 11$, 
value extrapolated for the number of light 
quark flavors being $n_f=2.5$ at the 
hadronization point.~\cite{Ham00} 
We obtain, for the fireball formed in Pb--Pb reactions,
$$0.2 \cdot  {11} T_{\mbox{\scriptsize h}}^4\simeq 0.17\mbox{ GeV/fm}^3\le {\cal B}\,.$$

\section{Thermal freeze-out}\label{thermal}
\subsection{Direct and decay $m_\bot$-spectra}
We next study the shape of  hadron $m_\bot$-spectra.
These are strongly influenced by 
resonance decays, which have also contributed in an important 
fashion to the yield of all stable hadrons observed.

The final particle distribution is composed of directly 
produced particles and decay products of heavier hadronic resonances:
\begin{equation}\label{2body}
\frac{dN_X}{dm_\bot} =\left.  \frac{dN_X}{dm_\bot}\right|_{\scriptsize\rm direct} +
\sum_{ \forall R \rightarrow X + 2+\cdots } 
\left.\frac{dN_X}{dm_\bot}\right|_{R \rightarrow X + 2 +\cdots}  
\end{equation}
Here,
$R(M,M_T,Y) \rightarrow X(m,m_T,y)+2(m_2)+\cdots$, where we 
indicate by the arguments that only for the decay  particle  $X$ 
we keep the information about the shape of the
momentum spectrum. 

In detail, the decay contribution to yield of 
$X$ is:
\begin{eqnarray}
\label{reso}
\frac{dN_X}{d {m^2_\bot} d y }&=&
\frac{g_{r} b}{4 \pi p^{*}}
\int_{Y_-}^{Y_+}\!\! dY
\int_{M_{T_-}}^{M_{T_+}} dM_{T}^{2}\, J \,
\frac{d^2 N_{R}}{dM_{T}^{2} dY}  
\\[0.3cm]\nonumber
J&=&\frac{M}{\sqrt{P_{T}^2 p_{T}^2 -\{M E^{*} - M_{T} m_{T} \cosh\Delta Y\}^2}}
\end{eqnarray}
We have used $\Delta Y=Y-y$, and
$\sqrt{s}$ is the combined invariant mass of the 
decay products other than particle $X$
and $E^{*}=(M^2-m^2-m_2^2)/2M$, $p^{*}=\sqrt{E^{*2}-m^2}$ 
are the energy and momentum of the decay
particle X in the rest frame of its parent.
The limits on the integration are the maximum values accessible
to the decay product $X$:
\[\ Y_{\pm}=y \pm \sinh^{-1}\left(\frac{p^{*}}{m_{T}}\right) \]
\[\ M_{T_{\pm}}=M 
\frac{E^{*} m_{T} \cosh\Delta Y \pm p_{T} 
\sqrt{p^{*2}-m_{T}^{2} \sinh^{2} \Delta Y}}
{m_{T}^{2} \sinh^{2} \Delta Y+m^{2}}\]

The theoretical primary particle
spectra (both those directly produced and parents of 
decay products) are derived from the Boltzmann 
distribution by Lorenz-transforming from a flowing 
intrinsic fluid element to the CM-frame, and 
integrating over allowed 
angles between particle direction  and local flow.

We introduce in the current analysis 
two velocities: a local  flow velocity $v$ of fireball 
matter where from particles emerge,
and hadronization surface (breakup) velocity which we refer to 
as $v_f^{\,-1}\equiv dt_f/dx_f$.  Particle production is controlled by the 
effective volume element, which comprises this quantity. In detail: 
\begin{equation}\label{v2v}
d S_{\mu} p^{\mu} = 
  d \omega \left(1- \frac{\vec v_{f}^{\,-1} \cdot \vec p}{E}\right)\,,
\qquad  d\omega \equiv \frac{d^3xd^3p}{(2\pi)^3}\,.
\end{equation}
The  Boltzmann distribution we adapt has thus the form
\begin{equation}
\frac{d^2 N}{dm_{T} dy} \propto
\left(1- \frac{\vec v_{f}^{\,-1} \cdot \vec p}{E}\right)
\gamma\,  m_{T} \cosh y \,
e^{-\gamma \frac E T \left(1-\frac{\vec v\cdot \vec p}{E}\right)}\,,
\end{equation}
where $\gamma=1/\sqrt{1-v^2}$\,.

The normalization for each hadron type $h=X,R$ is
\[\ N^h = V_{\scriptsize\rm QGP} \prod^{n}_{i\in h} \lambda_{i} \gamma_{i}\,. \]
We use the chemical parameters 
$\lambda_{i}, \gamma_{i}$ $i=q,s$ which  are as defined in~\cite{Let00}
and commonly used to characterize relative and absolute abundances of 
light and strange quarks.

The best thermal and chemical  parameters which minimize the total relative
error $\chi^2_{\rm T}$,
$$\chi^2_{\rm T}=\sum_i\left(\frac{F^{\scriptsize\rm \,theory}_i-F_i}
                        {\Delta F_i}\right)^2\,,$$
for all experimental measurement points $F_i$  having
measurement error $\Delta F_i$
are determined by considering simultaneously all strange 
baryon and antibaryon 
results of experiment WA97,~\cite{Ant00} 
{\it i.e.,}  $\Lambda$, $\overline{\Lambda}$, $\Xi$, $\overline{\Xi}$,
$\Omega + \overline{\Omega}$, as well as $K^0_S$, except that we assign 
additional systematic error to these data (see below).   

\subsection{$\Lambda$-, $\overline{\Lambda}$-, $\Xi$-, 
$\overline{\Xi}$-$m_\bot$-spectra}
In recent months experiment WA97 
determined the absolute normalization of the published 
$m_\bot$ distribution,~\cite{Ant00} and we
took the opportunity to perform the shape analysis
together with yield analysis in order to check if the 
thermal and chemical freeze-out conditions are the same.~\cite{Tor00}
Our analysis addresses the data differentiated in four different 
collision centralities. This allows to study the effect of the 
changing volume within the limits of precision set by the 
current data. 

Since particle spectra we consider have a good relative
normalization, only one parameter is required for 
each centrality in order to describe the absolute normalization
of all six hadron spectra. This is for two reasons important:\\
a) 
we can check if the volume from which strange hadrons 
are emitted grows with centrality of the collision as
we expect;\\
b) 
we can determine which region in $m_\bot$ 
produces the excess of $\Omega$  noted in the
chemical fit. \cite{Let00}

However, since the normalization $V_{\scriptsize\rm QGP}$ common for
all particles at given centrality comprises additional  
experimental acceptance normalization which we have not yet 
studied, we have  not normalized 
the value of the fireball emission volume at each centrality. 
Hence we will be presenting the volume parameter as
function of centrality  in arbitrary units.

If thermal and chemical freeze-outs are identical, our
present results  must  be consistent with 
earlier chemical analysis of hadron yields. Since
the experimental data we here study 
is dominated by the shape of $m_\bot$-spectra
and not by relative particle yields, our analysis is de facto 
comparing thermal and chemical freeze-outs. 

\begin{figure}[tb]
\vspace*{-1.1cm}
\centerline{\hskip 0.5cm
\epsfig{width=10.cm,clip=,angle=-90,figure=\pathnow 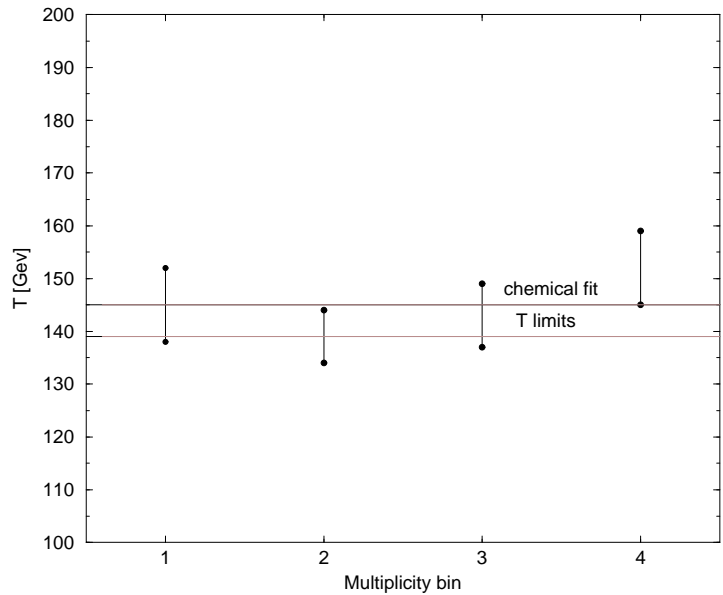}
}
\vspace*{-0.4cm}
\caption{ 
Thermal freeze-out temperature $T$ 
for different centrality bins compared to 
the chemical freeze-out analysis.
\label{TdT}
}
\end{figure}

We next show  the parameters
determining the shape of the $m_\bot$ distributions,
that is $T,v,v_f$.
As function of the centrality bin 1, 2, 3, 4 with the most
central bin being 4 we show in figure \ref{TdT}
the freeze-out temperature $T$ 
of the $m_\bot$ spectra. The horizontal lines 
delineate range of result of the chemical analysis.
It is reassuring that we find a result
consistent with the purely chemical analysis of data
that included non-strange hadrons.~\cite{Let00} 
There is 
no indication of a significant or systematic change of $T$ with centrality
though there is a somewhat higher value in the most central 4th bin. 
This is consistent with the believe that the formation of the new state of 
matter at CERN is occurring in all centrality bins explored by the 
experiment WA97. Only most peripheral interactions  
produce a change in the pattern of strange hadron production.~\cite{Kab99} 

The (unweighted) average of all results shown in figure \ref{TdT}
produces a freeze-out temperature at the upper boundary of the 
the pure chemical freeze-out
analysis result, $T\simeq 145$\,MeV. It should be noted that in chemical analysis we have not distinguished between the flow $v$ and freeze-out 
$v_f$ velocity and both are denoted as $v_c$,  which may be the cause of this 
slight difference between current analysis average and the earlier 
purely chemical analysis result. 

\begin{figure}[tb]
\vspace*{-1.1cm}
\centerline{\hspace*{1.0cm}
\epsfig{width=12cm,clip=,figure=\pathnow 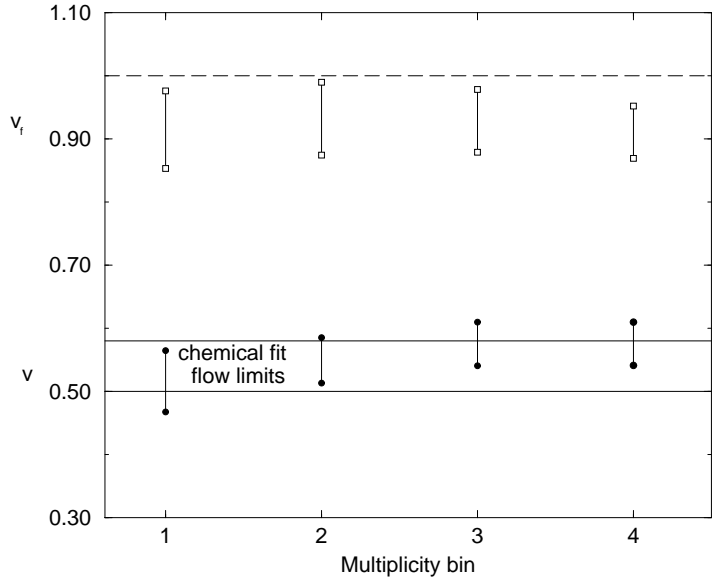}
}
\vspace*{-0.4cm}
\caption{ 
Thermal freeze-out flow velocity $v$ (top) and 
break up (hadronization) velocity $v_f$ 
for different centrality bins. Upper limit $v_f=1$ (dashed line) and 
chemical freeze-out analysis limits for $v$ (solid lines) are also shown.
\label{Tdv1v2}
}
\end{figure}

The magnitudes of the 
collective expansion velocity $v$  and the  break-up (hadronization) speed 
parameter $v_f$ are presented in figure \ref{Tdv1v2}.
For $v$ (lower part of the figure) 
we again see consistency with earlier chemical freeze-out
analysis results, and there is no confirmed systematic trend
in the behavior of this parameter as function of centrality. 

Though within the experimental error, one could argue 
inspecting  figure \ref{Tdv1v2} that there is 
systematic increase in transverse flow velocity $v$ with centrality and thus 
size of the system. This is expected, since the more central events comprise 
greater volume of matter, which allows more time for development
of the flow.  Interestingly, it is in $v$ and not $T$ that we find the 
slight change of spectral slopes noted in the presentation of the 
experimental data.~\cite{Ant00} 

The value of the beak-up (hadronization) speed 
parameter $v_f$ shown in the top portion of figure \ref{Tdv1v2} is near to 
velocity of light which is consistent with the picture of a 
sudden breakup of the  fireball. This 
hadronization surface velocity $v_f$ was in the earlier chemical
fit set to be equal to $v$, as there was not enough sensitivity in
purely chemical fit to  determine the value of $v_f$. 

\begin{figure}[tb]
\vspace*{-1.1cm}
\centerline{\hspace*{0.5cm}
\epsfig{width=9.cm,clip=,angle=-90,figure=\pathnow 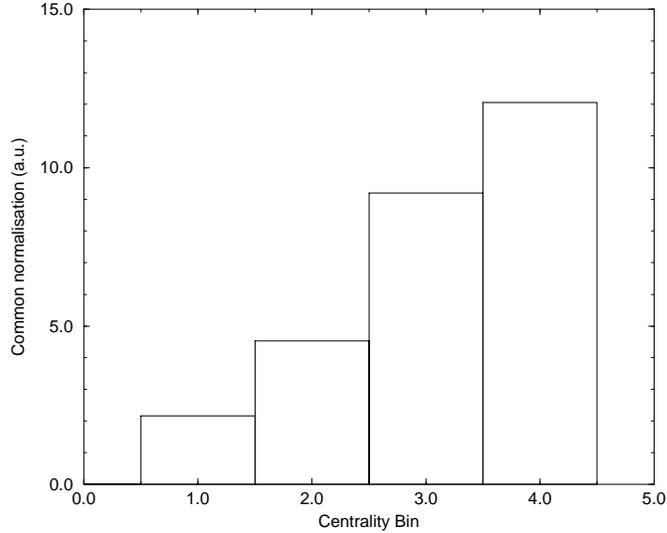}
}
\vspace*{-0.2cm}
\caption{ 
Hadronization volume (arbitrary units)
 for different centrality bins. 
\label{Tdiagnorm}
}
\end{figure}

Unlike the temperature and two velocities, the overall normalization
of hadron yields, $V^h$ is found to be strongly centrality dependent, as is
seen in figure \ref{Tdiagnorm}. This confirms in quantitative way the
believe that the entire available fireball volume is available for 
hadron production. The strong increase in the volume
by factor six is qualitatively 
consistent with a geometric interpretation of the 
collision centrality effect. Not shown is the error propagating from the
experimental data which is strongly correlated to the chemical
parameters discussed next. This systematic uncertainty is another reason 
we do not attempt an absolute unit volume normalization. 

The four chemical parameters
 $\lambda_q,\lambda_s, \gamma_q, \gamma_s/\gamma_q$ are
shown in the following figures \ref{Tdlqls}, \ref{Tdgamqgams}.
These parameters determine along with $V^h$ the final particle
yield. Since we have 5 parameters 
determining normalization of  6 strange hadron spectra, and as discussed we 
reduce the statistical weight of Kaons, there
is obviously a lot of correlation between these 4 quantities, and thus 
the error bar which reflects this correlation,  is significant.

\begin{figure}[tb]
\vspace*{-2.3cm}
\centerline{\hskip -2.6cm
\epsfig{width=10.5cm,clip=,angle=-90,figure=\pathnow 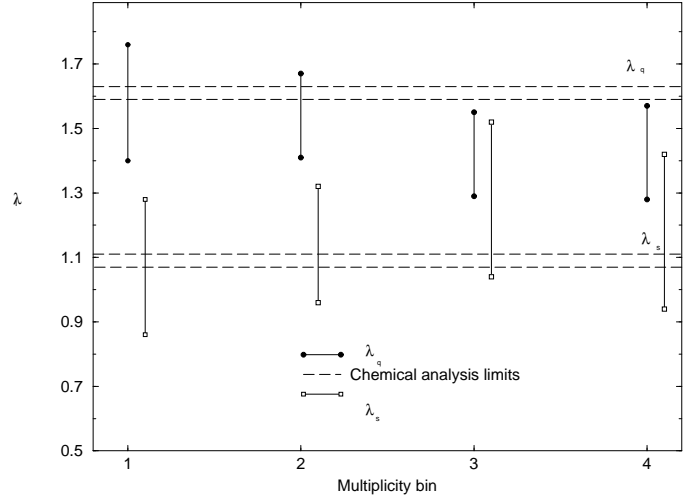}
}
\vspace*{-0.2cm}
\caption{ 
Thermal analysis chemical quark fugacity $\lambda_q$ (top) and 
strange quark fugacity $\lambda_s$ (bottom)
 for different centrality bins compared to 
the chemical freeze-out analysis results. 
\label{Tdlqls}
}
\end{figure}
The chemical fugacities $\lambda_q$ and $\lambda_s$ shown in 
figure \ref{Tdlqls} do not exhibit
a systematic centrality dependence. This is consistent with the result
we found for $T$ in that the freeze-out properties of the fireball are
seen to be for the temperature and chemical potential values independent
of the size of the fireball. Comparing to the earlier chemical
freeze-out result in figure \ref{Tdlqls} one may argue that 
there is a systematic downward deviation in $\lambda_q$. However,
 this could be  caused by the fact that the 
chemical freeze-out analysis allowed for
isospin-asymmetric $\Xi^-(dss)$ yield,~\cite{Let00} 
while out present analysis is not
yet distinguishing light quarks.
 
The ratio $\gamma_s/\gamma_q$  shown in bottom portion of  figure \ref{Tdgamqgams}
 is systematically smaller than unity, consistent with many years of prior analysis:
when $\gamma_q=1$ is tacitly chosen, this ratio is the value of $\gamma_s$
in analysis of strange baryons.
We have not imposed a constrain on the range of $\gamma_q$ 
(top of figure \ref{Tdgamqgams}) and thus 
values greater than the pion condensation point 
$\gamma_q^*=e^{m_\pi/2T}\simeq 1.65$ (thick line) can be expected, but in fact 
do not arise.

\begin{figure}[tb]
\vspace*{-2.5cm}
\centerline{\hskip -1.cm
\epsfig{width=9.cm,height=10.cm,clip=,angle=-90,figure=\pathnow 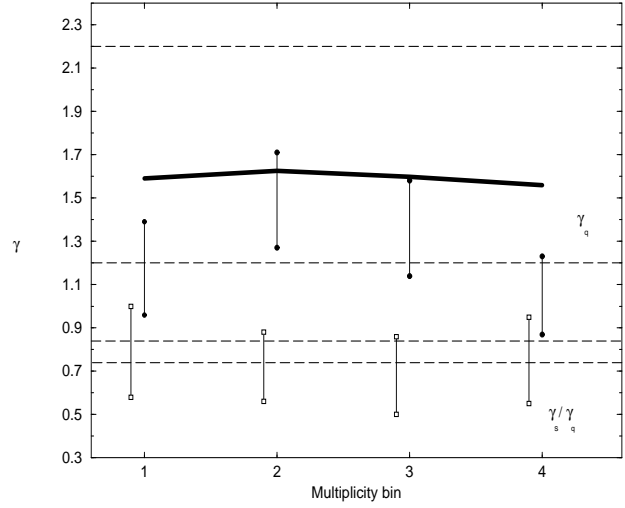}
}
\caption{ 
Thermal analysis chemical quark abundance parameter $\gamma_q$ (top)
and $\gamma_s/\gamma_q$ (bottom)  for different centrality bins compared to 
the chemical freeze-out analysis. Thick line: upper limit arising with pion
condensation.
\label{Tdgamqgams}
}
\end{figure}
\begin{figure}[tb]
\vspace*{-1.cm}
\centerline{
\epsfig{width=8.5cm,clip=,angle=-90,figure=\pathnow 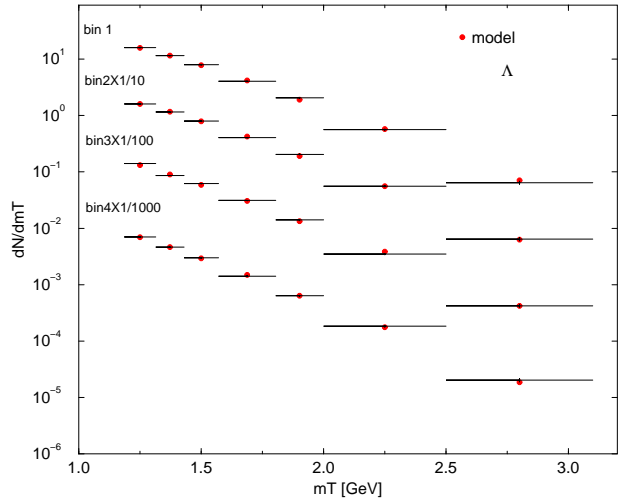}
}
\vspace*{-0.6cm}
\caption{ 
Thermal analysis $m_T$ spectra: $\Lambda$ 
\label{TdLamtot}
}
\end{figure}

\begin{figure}[tb]
\vspace*{-1.cm}
\centerline{
\epsfig{width=8.5cm,clip=,angle=-90,figure=\pathnow 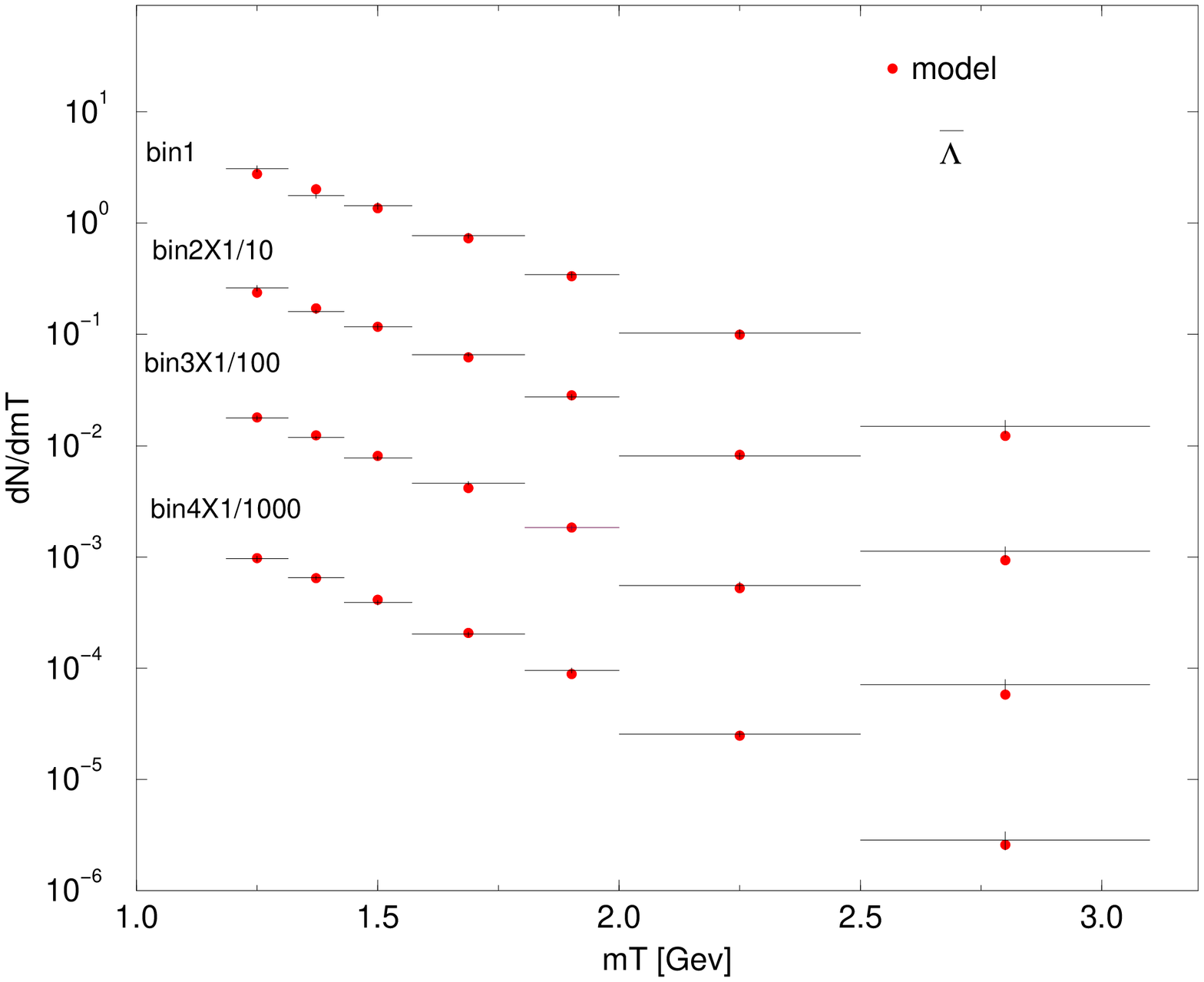}
}
\vspace*{-0.6cm}
\caption{ 
Thermal analysis $m_T$ spectra: $\overline\Lambda$. 
\label{TdALamtot}
}
\end{figure}

\begin{figure}[tb]
\vspace*{-1.cm}
\centerline{
\epsfig{width=8.5cm,clip=,angle=-90,figure=\pathnow 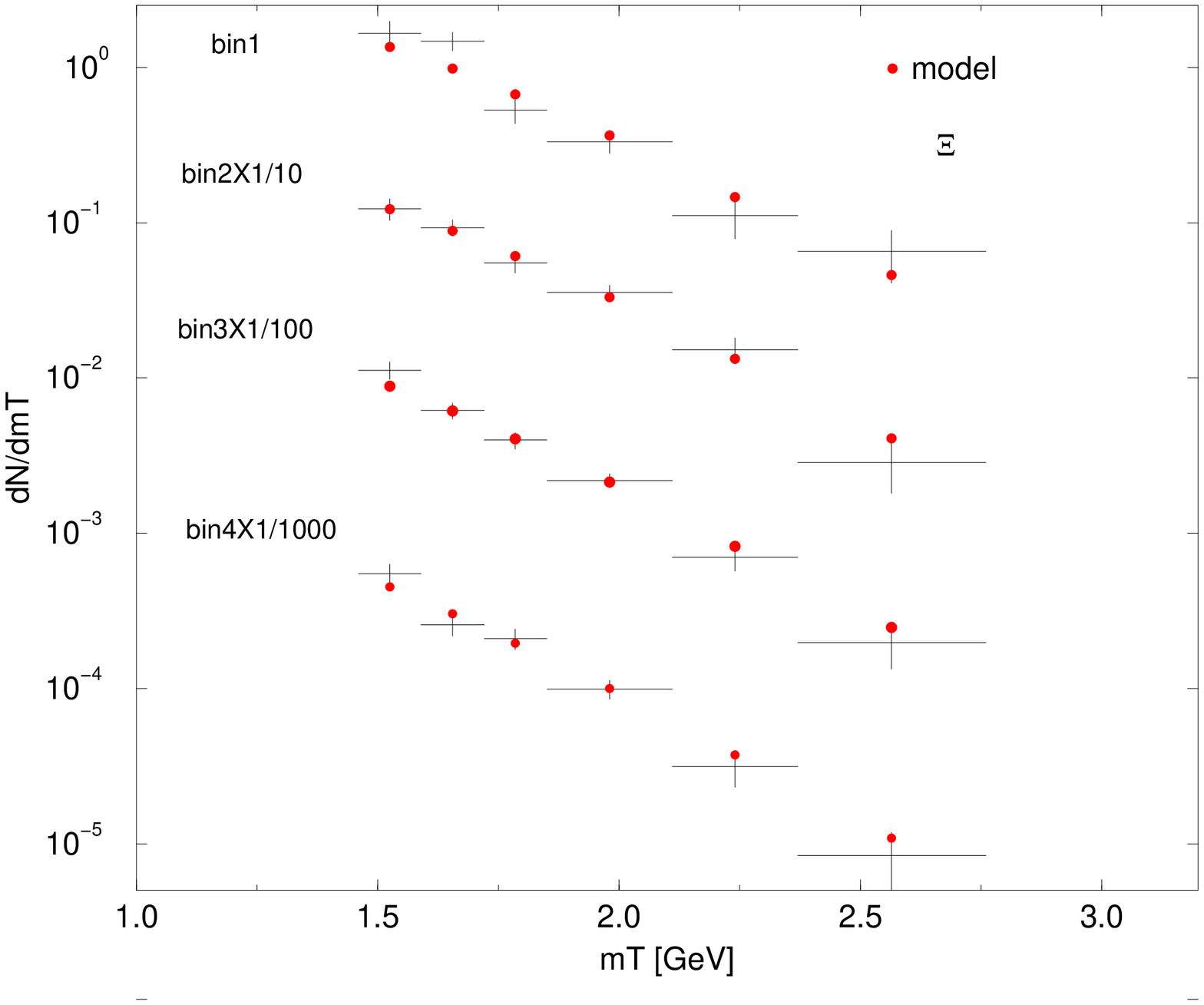}
}
\vspace*{-0.6cm}
\caption{ 
Thermal analysis $m_T$ spectra: $\Xi$
\label{TdXitot}
}
\end{figure}

\begin{figure}[tb]
\vspace*{-1.cm}
\centerline{
\epsfig{width=8.5cm,clip=,angle=-90,figure=\pathnow 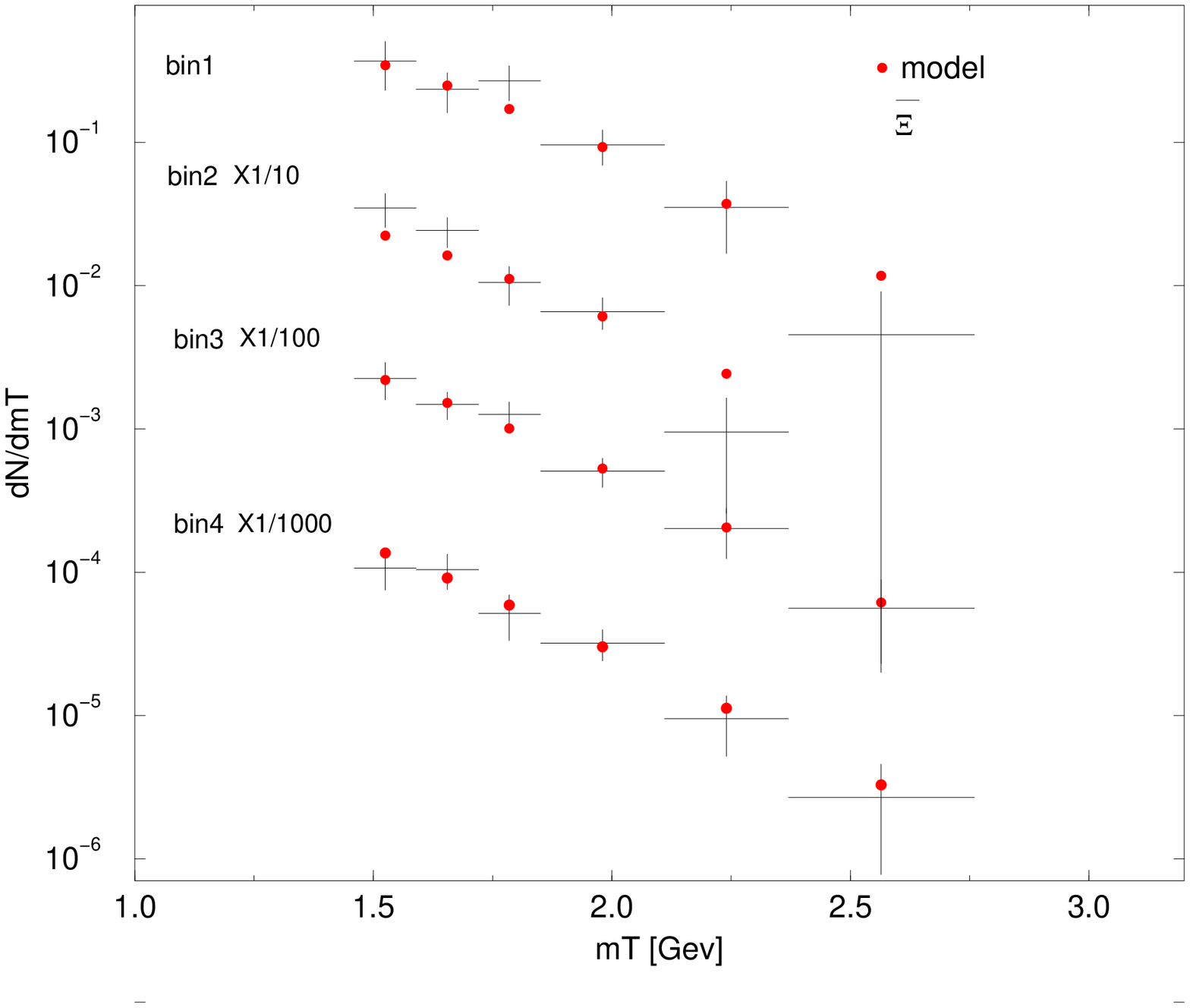}
}
\vspace*{-0.6cm}
\caption{ 
Thermal analysis $m_T$ spectra: $\overline\Xi$
\label{TdAXitot}
}
\end{figure}

It is important to explicitly check
how well the particle $m_\bot$-spectra are
reproduced. We group all bins in one figure and show 
in figures \ref{TdLamtot},
\ref{TdALamtot}, \ref{TdXitot}, \ref{TdAXitot} 
in sequence  $\Lambda,\,\overline\Lambda,\,\Xi,\,\overline\Xi$. 
Overall, the description of the shape of the spectra 
is very satisfactory.

\subsection{Omega and Meson spectra}
\begin{figure}[tb]
\vspace*{-1.cm}
\centerline{\hskip 0.5cm
\epsfig{width=8.5cm,clip=,angle=-90,figure=\pathnow 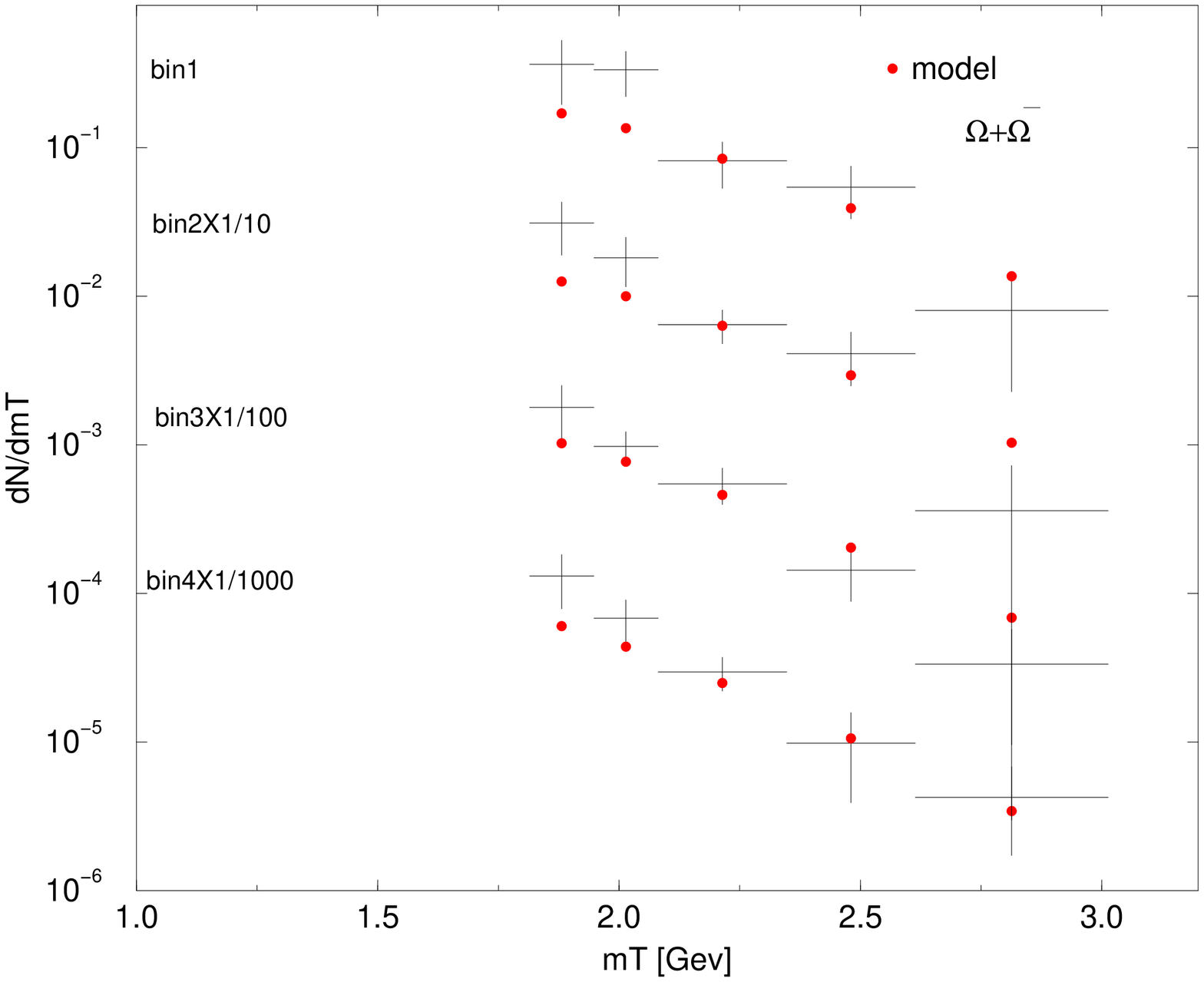}
}
\vspace*{-0.6cm}
\caption{ 
Thermal analysis $m_T$ spectra: $\Omega+\overline\Omega$.
\label{TdOAOtot}
}
\end{figure}

In figure \ref{TdOAOtot}
all four centrality bins for the sum $\Omega+\overline\Omega$ are
shown. We see that we systematically  under predict the two lowest
$m_\bot$ data points. Some deviation at high $m_\bot$ may be attributable
to acceptance uncertainties, also seen in the the 
$\Xi$ result presented earlier in figure \ref{TdAXitot}. 
 We recall that there is a disagreement between theoritical 
expectation and the 
Omega yields. 
In the here presented thermal
analysis, we see that this disagreement is arising at
low momentum. We could include all experimental results in the
present analysis since the omega data have a relative low statistical
evidence and thus other hyperon data determine
the behavior of $\Omega$-spectrum. 

The low-$m_\bot$ anomaly also explains why the 
inverse $m_\bot$ slopes for $\Omega,\,\overline\Omega$ 
are smaller than the values seen in all other strange (anti)hyperons.
One can presently only speculate about the processes 
which contribute to this anomaly. 
We note that the 1--2s.d. deviations in the 
low  $m_\bot$-bins of the $\Omega+\overline\Omega$ 
spectrum translates into 3s.d. 
deviations from the prediction of the chemical analysis. 
This is mainly a consequence of the fact that after summing
over centrality and $m_\bot$, the statistical error which dominates
$\Omega+\overline\Omega$ spectra becomes relatively small, and
as can be seen the low $m_\bot$ excess practically doubles
the yield. 

For kaons K$^{0}$ the statistical errors are
very small, and we find in a more in depth statistical analysis that
they must be smaller than the systematic errors not considered. 
This is done checking the influence of the deletion of individual 
measurement points  in the spectrum on the fit result, 
which influences the outcome  in case of kaons only. We 
 assign in a somewhat arbitrary manner a systematic error 
that we arbitrarily choose to be 5 times greater than statistical error.

\begin{figure}[tb]
\vspace*{-1.cm}
\centerline{
\epsfig{width=8.5cm,clip=,angle=-90,figure=\pathnow 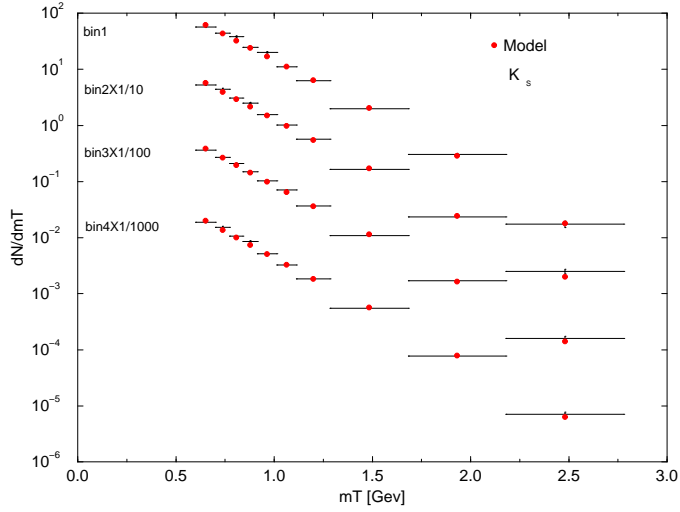}
}
\vspace*{-0.6cm}
\caption{ 
Thermal analysis $m_T$ spectra: $K_s$.
\label{TdK0All}
}
\end{figure}

With this modification we included the kaon data in the analysis
which included the  above presented strange baryon and antibaryon
results. Because of the large error we assigned, there is relatively little
influence the kaon spectra have on the baryon/antibaryon analysis. In fact
the kaon results show how the hyperon/antihyperon 
experimental results predict K$^0_S$-$m_\bot$ spectra.
We  describe the kaon data well, and especially so 
in the $m_\bot$ range which is the same as that for
hyperons considered earlier,  as is seen in figure \ref{TdK0All}.

We now turn to the pion spectra measured by the WA98 
collaboration.~\cite{WA98pi0}  This is a very difficult 
matter, since at low momenta there are many
mechanisms of pion production which we have not analyzed. 
Moreover, we have not fully understood how the longitudinal
flow influences through the secondary decays the observed 
pion spectra. Compared to strange hadrons
 we study the $\pi^0$ data in a very wide
range of $m_\bot$ in which the yield changes
by 6 orders of magnitude. 

\begin{figure}[tb]
\vspace*{-1.3cm}
\centerline{
\hspace*{0.cm}\epsfig{width=9.cm,clip=,angle=-90,figure=\pathnow 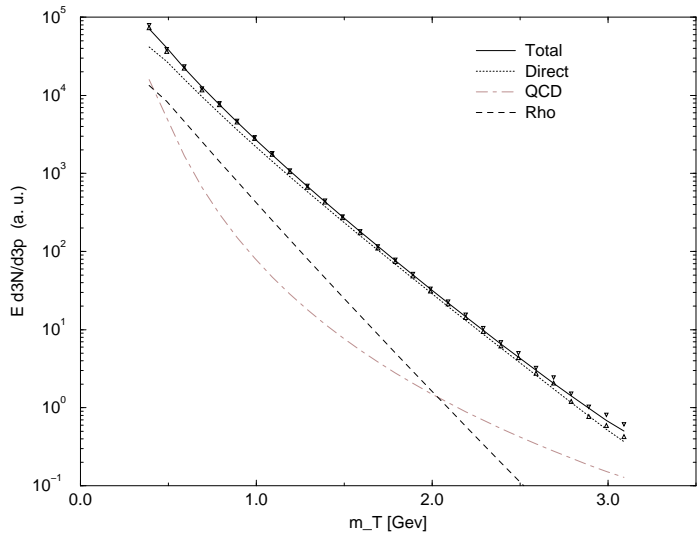}
}
\vspace*{-0.6cm}
\caption{ 
Central rapidity data of experiment WA98 for $\pi^0$ are 
compared to the spectra expected in a sudden freeze-out reaction picture.
See text for details.
\label{Ppi01}
}
\end{figure}

We use a set of parameters obtained fro the hyperon data, {\it i.e.,} $T_f=143$\,MeV, setting also  freeze-out surface velocity $0.99c$. 
We  include in the results shown
in figure \ref{Ppi01} aside of  directly produced pions the two body decay of 
the $\rho$. We  allow 
a direct hard parton QCD component contribution,~\cite{Fey77}
of the form $$Ed^3N/d^3p\propto 1/p_\bot^\kappa\,.$$ 
Thus, we vary four parameters:
two normalizations (thermal and QCD direct component), and also 
$\kappa= 5.6\pm1.2$ , $v_c=0.55\pm0.02$, with resulting best fits as 
stated. This procedure 
yields a good description of the data as seen 
in figure \ref{Ppi01}, with $\chi^2/\mbox{dof}<1.4$. Considering that only
statistical error is being  considered, and we were not evaluating 
contributions of 3-body decay resonances, we see this as a noteworthy
confirmation that the hard pions are also created primarily
in the sudden hadronization process.  We note that the direct QCD component
 is at 1--20\% level, and thus before proceeding further 
a more  precise model of this contribution needs to be developed. 

Our analysis is not contradicting results shown 
in~\cite{WA98pi0}, for these
authors did not consider that the freeze-out 
surface velocity is different from 
the flow velocity, and they  did not allow 
for direct parton-parton scattering
contributions in their analysis of pion spectra.  

Our results confirm 
the finding that the apparent temperature hierarchy for different mass
particles  is due to a collective
expansion of the source.~\cite{NuX98}

\section{Strangeness at RHIC}\label{RHIC}
\subsection{Expectations}

The abundances of particles produced from QGP  
within the sudden  freeze-out model are controlled by several 
parameters we restate here: the light quark fugacity 
$1<\lambda_{q} <1.1$\,,  value is limited by the expected small
ratio between baryons and mesons (baryon-poor plasma) when the 
energy per baryon is above 100\,GeV, strangeness fugacity 
$ \lambda_{s}\simeq 1$ 
which value for locally neutral plasma assures that 
$\langle s-\bar s\rangle =0$; 
the light quark phase space occupancy 
$\gamma_{q}\simeq 1.5$, overabundance value due to  gluon fragmentation. 
Given these narrow ranges of chemical parameters and 
the sudden freeze-out temperature $T_f=150$ MeV
expected for nearly baryon free domain, 
 we compute the particle production at break-up. 

Taking $\gamma_{q}=1.5\!\!{\tiny\begin{array}{c}+0.10\\-0.25 \end{array}}\!\!$\ , 
we choose the value of 
$\lambda_{q}$, see the header  of table \ref{table1}, 
for which the  energy  per baryon ($E/B$)
is similar to the collision condition 
(100\,GeV),  which leads to the 
range $\lambda_{q}=1.03\pm0.005$. We  evaluate for 
these  examples aside of $E/B$, the strangeness per baryon 
$s/B$ and entropy per baryon $S/B$ as shown in the top 
section of the table \ref{table1}. We do
not enforce  $\langle s-\bar s\rangle=0$ 
at central rapidity exactly, but 
since baryon asymmetry is
small, strangeness is  balanced to better than 2\%\, in
the parameter range considered.
In the bottom portion of  table \ref{table1},  we present
the compatible particle abundance ratios,
computed according to the procedure  we developed,
and described above.~\cite{Let00} We have given,
aside of the baryon and antibaryon relative yields, also the relative 
kaon yield, which is also well determined within this approach. 

\begin{table}[t]
\caption{\label{table1} 
For $\gamma_{s}=1.25,\,\lambda_{s}=1$ and $\gamma_{q}$, $\lambda_{q}$ as shown:
Top portion: strangeness per baryon $s/B$, 
energy per baryon $E/B$[GeV]  and  entropy per baryon $S/B$. Bottom portion:
sample of hadron ratios expected at RHIC.}
\small
\vspace*{-0.2cm}
\begin{center}
\begin{tabular}{llllll}\\
\hline
 $\gamma_{q}$                               & 1.25 & 1.5  &  1.5  &  1.5  & 1.60 \\
$\lambda_{q}$                               & 1.03 & 1.025&  1.03 & 1.035 & 1.03 \\
\hline
$E/B$[{\small GeV}]\ \                        & 117  & 133  &  111  &  95   & 110 \\
$s/B$                                     & 18   & 16   &  13   &  12   & 12 \\
$S/B$                                     & 630  & 698  &  583  & 501   & 571 \\
\hline
$p/{\bar p}$                              & 1.19 & 1.15 & 1.19  &  1.22 & 1.19 \\
$\Lambda/p$                               & 1.74 & 1.47 & 1.47  &  1.45 & 1.35 \\
${\bar\Lambda}/{\bar p}$                  & 1.85 & 1.54 & 1.55  &  1.55 &1.44 \\
${\bar\Lambda}/{\Lambda}$                 & 0.89 & 0.91 &  0.89 &  0.87 & 0.89 \\
${\Xi^-}/{\Lambda}$                         & 0.19 & 0.16&  0.16  &  0.16 & 0.15 \\
${\overline{\Xi^-}}/{\bar\Lambda}$          & 0.20 & 0.17 &  0.17 &  0.17 & 0.16 \\
${\overline{\Xi}}/{\Xi}$                  & 0.94 & 0.95 &  0.94 &  0.93 & 0.94 \\
${\Omega}/{\Xi^-}$                                    & 0.147&0.123 &  0.122&  0.122& 0.115 \\
${\overline{\Omega}}/{\overline{\Xi^-}}$              & 0.156& 0.130&  0.130&  0.131& 0.122 \\
${\overline{\Omega}}/{\Omega}$                      &  1   & 1.   &  1.   &  1.   & 1.   \\
$\Omega+\overline{\Omega}\over\Xi^-+\overline{\Xi^-}$    & 0.15 & 0.13 &  0.13 &  0.13 & 0.12 \\
$\Xi^-+\overline{\Xi^-}\over\Lambda+\bar{\Lambda}$       & 0.19 & 0.16 &  0.16 &  0.16 & 0.15 \\
${K^+}/{K^-}$                                       & 1.05 & 1.04 &  1.05 &  1.06 & 1.05 \\
\hline\vspace*{-0.6cm}
\end{tabular}
\end{center}
\end{table}

The meaning of these results can be better appreciated when
we assume in an example the central  rapidity density
of direct protons is  $dp/dy|_{\mbox{\scriptsize cent.}}=25$. 
In table  \ref{table2},  we present the 
resulting (anti)baryon abundances. 
The  net baryon density $db/dy\simeq 16\pm3$,  there is 
baryon number transparency. We see that (anti)hyperons are 
indeed more abundant than non-strange  (anti)baryons.
Taking into account the disintegration of strange baryons,
we are finding a much greater  number of observed protons 
$dp/dy|_{\mbox{\scriptsize cent.}}^{\mbox{\scriptsize obs.}}
\simeq 65\pm5$ in the central rapidity region. It is important
when quoting results from table  \ref{table2} to recall that:\\
\indent 1) we have chosen arbitrarily the overall
normalization in table  \ref{table2}\,, only particle ratios 
were computed,  and\\ 
\indent 2) the rapidity baryon density
relation to rapidity proton density is a consequence of the assumed 
value of $\lambda_{q}$, which we chose to get 
$E/B\simeq 100$\,GeV per  participant. 

\begin{table}[tb]
\caption{\label{table2} $dN/dy|_{\mbox{\scriptsize cent.}}$ 
assuming in this example $dp/dy|_{\mbox{\scriptsize cent.}}=25$ .}
\vspace*{-.2cm} 
\begin{center}
\begin{tabular}{ll|cccccccccc}\\
\hline
$\gamma_{q}$& $\lambda_{q}$  & $b$ &  $p$ & $\bar p$ & $\!\!\Lambda\!\!+\!\!\Sigma^0\!\!$ & $\!\!\overline{\Lambda}\!\!+\!\!\overline{\Sigma}^0\!\!$&$\Sigma^{\pm}$&$\overline{\Sigma}^{\mp}$ & 
$\Xi^{^{\underline{0}}}$ &$\overline{\Xi}^{^{\underline{0}}}$& $\Omega\!=\!\overline\Omega$  \\
\hline
1.25& 1.03 & 17 & 25$^*$& 21 & 44 & 39 & 31 & 27 & 17 & 16 & 1.2  \\ 
1.5 & 1.025 & 13 & 25$^*$& 22 & 36 & 33 & 26 & 23 & 13 & 11 & 0.7 \\ 
1.5& 1.03 & 16 & 25$^*$& 21 & 37 & 33 & 26 & 23 & 12 & 11 & 0.7 \\ 
1.5 & 1.035 & 18 & 25$^*$& 21 & 36 & 32 & 26 & 22 & 11 & 10 & 0.7 \\ 
1.60& 1.03  & 15 & 25$^*$& 21 & 34 & 30 &24 &21 & 10 & 9.6 & 0.6 \\ 
\hline\vspace*{-0.6cm}
\end{tabular}
\end{center}
\end{table}

\subsection{Comparison of theory and experiment}
Our study of RHIC results has just begun and we expect that
it remains an ongoing project as we enter the period
of RHIC data collection and analysis. Some of our predictions 
can already be compared with experiment (STAR at RHIC).
These results are in agreement with our model of how 
QGP evolves at RHIC.

In order to compare with experiment, we need to adjust 
from the assumed 100+100$A$\,GeV RHIC energy 
to the RHIC run energy in 2000 
which is  65+65$A$\,GeV. We consider the  $\overline\Lambda/\Lambda$
ratio.  At 8.6+8.6$A$\,GeV (corresponding to 160$A$\,GeV 
the experimental result is $\overline\Lambda/\Lambda\simeq 0.12$
while at 100+100$A$\,GeV we predict 
$\overline\Lambda/\Lambda = 0.89\pm0.02$
in  table  \ref{table1}. We also made a prediction  for 35+35$A$\,GeV~\cite{Bas00},   where we found
$\overline\Lambda/\Lambda\simeq 0.5$. The interpolated result 
for the  65+65$A$\,GeV RHIC run is 
$\overline\Lambda/\Lambda\simeq 0.68\pm0.05$,
which is nearly exactly the October 2000 APS-meeting
STAR ratio $\overline\Lambda/\Lambda = 0.7 \pm 0.05 \pm 0.2$\,. 
From our $\overline\Lambda/\Lambda$ ratio one obtains 
$$\frac{\bar p}{p}= \frac{\lambda_s^2}{\lambda_q^2}\frac{\overline\Lambda}{\Lambda}\,, $$
and this implies taking note of the heading and caption
in  table  \ref{table1} that the $\bar p/p$ ratio is about 7\% smaller
compared to  $\overline\Lambda/\Lambda$ ratio. In fact  STAR reported 
$\bar p/p=0.65\pm0.05$, in agreement with our result.

Both particle ratios can thus be well described within  our 
understanding of the hadronization and properties of QGP at RHIC.
In their absolute values reflect on the expected small baryon content
at the initial RHIC energy. Both should increase by 25--30\% at the
top RHIC energy (100+100 $A$GeV).

The most interesting result seen in table  \ref{table2}\,, 
which confirmation we anticipate and await with anxiety is 
the hyperon-dominance of the baryon 
yields at RHIC, a fact which as we believe does not depend on a detailed 
model hypothesis.  Another interesting property of the hadronizing hot RHIC
matter, as seen in  table \ref{table1}, is that strangeness yield
per participant is expected to be 13--23 times 
greater than seen at present in Pb--Pb interactions at
SPS energies, where we have 0.75 strange quark pairs per baryon.

\section{Summary and conclusions}\label{final}
The enhanced strangeness yield observed at 158--200$A$ GeV reactions
corresponds according to our study  to  $\cal O$(0.75) $s\bar s$-pairs
of quarks per participant baryon, see section \ref{analyze}. 
Considering the properties of QGP-liquid explored in section \ref{QGP}
 and the resulting initial gluon temperatures we have shown 
that this exceptionally high yield 
is achievable in a short time that the collision is known to last 
by in-plasma gluon-fusion reactions, $G+G\to s+\bar s$, 
as was proposed many years ago.~\cite{Raf82} 

Beyond strangeness enhancement, multi
strange baryons and antibaryons have been recognized as being
even a more specific source of information about the deconfined 
state of the matter. The analysis of experimental
results from 158$A$GeV Pb--Pb interactions carried out in
sections \ref{analyze} (chemical abundance analysis) and in 
section \ref{thermal} (spectral $m_\bot$ shape) confirms this.
We find that strange baryons and antibaryons are 
dominantly produced in sudden hadronization of the quark-gluon
fireball with chemical and thermal decoupling occurring at the
same time. 

From the appearance of first results we recognized that sudden
hadronization is required.~\cite{Raf91} However, 
initially the mechanisms leading to direct hadronization 
(hadron production from QGP fireball without equal rescattering),
have not been theoretically understood. Here,  we
have presented in section \ref{instability} 
 a complete reaction picture consistent with
QGP equations of state.~\cite{Raf00} The
sudden breakup (direct hadronization) 
into final state  particles occurs as the QGP 
fireball super-cools, and in this 
state encounters a mechanical instability, which drives the explosive 
disintegration.
Deep supercooling requires a  first order phase transition, and
this in turn implies presence of a latent heat and we have estimated
its magnitude in section \ref{instability}. 

There are several experimental facts that lead us
to consider sudden QGP hadronization. The most recent 
point has been the results presented in section \ref{thermal}
which show agreement between thermal and chemical freeze-out
analysis. An important experimental input is the 
identity of the $m_\bot$ spectra for strange baryons and 
antibaryons. We have been motivated initially to pursue 
sudden QGP hadronization by  the need to use chemical
nonequilibrium to describe hadron production. 
At first we have proposed this for the case of strange quarks,
and more recently as the experimental data became
more comprehensive, we also recognized the need to 
consider light quark chemical nonequilibrium --- we described details
of these developments elsewhere.~\cite{Let00} 

The light quark pair excess is a convenient way to understand hadron 
multiplicity excess, which we have related 
to an entropy excess.~\cite{Let93} But is this the only good way to understand the
experimental hadron production data? It is hard to check all
possibilities. However, we explored (in medium) changes of hadron masses.
We have found that invariably the statistical significance of the analysis 
decreases as we modify individual hadron properties, as long as we have been
assuming  chemical equilibrium of hadron abundances, and a consistent
change of both mesons and baryons properties. Our extensive 
trial and error searches have convinced us that the only theoretical 
description of hadron production data that  works ($\chi^2/\mbox{dof}<1$) requires excess of valence light quark pair abundance, irrespective of the 
detailed strategy of data analysis.

So how it can be that
 there are many people who believe that one can describe
hadron production data assuming chemical equilibrium of
light quarks. In fact, if we take hadronization to somehow occur
in chemical equilibrium, or/and ignore the possibility of
plasma formation, we find $\chi^2/\mbox{dof}>2.5,\ \mbox{dof}=10$ 
in our analysis, which agrees when carried out with 
this assumption with 
results of others.~\cite{BM99} Even though this $\chi^2$
is without convincing physical significance, what has 
caused that this result has not been forgotten is  that
the chemical freeze-out  temperature for equilibrium condition 
comes out to be near to the expected equilibrium phase transition condition, 
$T\simeq 170$\,MeV, see section \ref{QGP}. Thus the assumption 
of equilibrium seems to be consistent with the properties of QGP
phase. Furthermore, the study of soft pions suggests
(unlike it is with hard pions or other hadrons) 
that the thermal freeze-out occurs
at about 120\,MeV. Naturally, soft pions can be produced by 
more complex collective processes 
not considered in the conventional thermal analysis and we 
thus did not introduce these particles into our study, preferring to
depend on the numerous other results in section \ref{thermal}.

Further study of details of chemical freeze-out, involving
consideration of strong resonances may lead to a direct 
experimental differentiation of these two alternatives. We
are presently considering $\Sigma^*(1385)$,~\cite{Tor01}
 which if freeze-out occurs at $T=170$\,MeV would be the 
dominant source of $\Lambda$ considering the decay 
$\Sigma^*(1385)\to \Lambda+\pi$ (88\%). But 
we are already today quite certain about the correctness of 
the sudden hadronization picture since:
\begin{enumerate}
\item the experimental strange and non-strange, 
hadron abundance results  statistically strongly favor
chemical nonequilibrium in hadronization; 
\item  baryons and antibaryons have the same $m_\bot$ spectra; and 
\item  our recent finding from the analysis of $m_\bot$ spectra
described in section~\ref{thermal}
that the  thermal freeze-out temperature is the same as the 
chemical  non-equilibrium freeze-out condition.
\end{enumerate}

Our enhanced understanding of the fireball supercooling and
resulting sudden breakup, as well as our study of thermal
freeze-out properties has in the past year considerably strengthened
our case. Not only we can now argue that the point of instability
agrees with the thermal and chemical freeze-out, but also 
we have considered  the freeze-out surface
dynamics and have shown that the break-up velocity $v_f$ is nearly
velocity of light, as would be expected in  a sudden breakup of a
QGP fireball. 

What are the possible causes of the chemical non-equilibrium?
Certainly even if the breakup of the QGP-fireball is sudden
we could have abundance equilibrium. In fact, the microscopic 
processes governing the fireball
breakup determine how the physical and statistical properties
of the fireball change at the breakup point. 
In particular as gluons convert into quark 
pairs and hadrons, the gluon abundance parameter
 $\gamma_g\to 0$ and  the quark 
chemical occupancy parameters have to increase significantly,
as we determine in data analysis.  On the other hand 
the energy $E$  and baryon content  $b$ of the fireball are  conserved. 
Entropy $S$ is conserved when the gluon content of a 
QGP fireball is transformed into 
quark pairs in the entropy conserving  
process $G+G\to q+\bar q$.  Similarly, when
quarks and antiquarks recombine into hadrons,  entropy is conserved in the 
range of parameters of interest here.  Thus  also $E/B$ and $S/B$ is conserved 
across hadronization condition. The sudden hadronization process 
also maintains the temperature $T$ and baryo-chemical potential $\mu_b$
across the phase boundary.

We have described strange hadron $m_\bot$-spectra with 
astounding precision in all available centrality bins. 
The remarkable similarity of 
$m_\bot$ spectra reported by  the  WA97 experiment
is interpreted by a set of freeze out parameters,
and we see that production mechanism of 
$\Lambda$, $\overline{\Lambda}$, and  $\Xi$, 
$\overline\Xi$ is the same. This symmetry, including matter--antimatter 
production is an important cornerstone of the claim that the strange
antibaryon data can only be interpreted in terms of direct particle emission
from a deconfined phase. In presence of conventional 
hadron collision based physics, the production mechanism 
of antibaryons is quite different from that of baryons 
and a similarity of the $m_\bot$ spectra is not 
expected. Moreover, even if QGP is formed, but a equal
phase of confined particles is present, the annihilation of 
antibaryons in the baryon rich medium created at CERN-SPS energy
would deplete more strongly antibaryon yields, in
particular so  at small momenta, 
with the more abundant baryons remaining less influenced. 
This effect is not observed.~\cite{Ant00} 

Similarity of $m_\bot$-spectra 
does not at all imply, and one should not expect 
a similarity of particle rapidity spectra.
As hyperon are formed at the fireball breakup, 
any remaining longitudinal
flow present among fireball constituents 
will be imposed on the product particle, thus $\Lambda$-spectra
 containing potentially two original valence quarks are stretched in
$y$, which  $\overline\Lambda$-$y$-spectra are not,
 as they are made from newly 
formed particles. All told, one would expect 
that anti-hyperons can appear with a thermal
rapidity distribution, but hyperons will not. 
But both have the same  thermal-explosive collective 
flow controlled shape of $m_\bot$-spectra.

The only systematic disagreement we see is that the 
 $\Omega+\overline\Omega$ results have
a noticeable systematic low $p_\bot$ enhancement anomaly visible
in two lowest $m_\bot$ bins in 
all centrality bins. This result shows that it is not a
different temperature of freeze-out  of $\Omega+\overline\Omega$ 
 that leads to more enhanced yield, but a soft momentum secondary source
which contributes almost equal number of soft $\Omega+\overline\Omega$
compared to the  yield expected in view of the behavior of other
strange hadrons.

We have presented here a short overview of our recent work 
developing further strangeness as a  signature 
of the formation of quark-gluon plasma, and as a 
diagnostic tool allowing the study of its properties. 
The understanding of the properties of the QGP formed 
in CERN-SPS experiment has greatly improved in past 18 months. 

\section*{Acknowledgments}
Work supported in part by a grant from the U.S. Department of
Energy,  DE-FG03-95ER40937\,.\\ Laboratoire de Physique Th\'eorique 
et Hautes Energies, LPTHE, at  University Paris 6 and 7 is supported 
by CNRS as Unit\'e Mixte de Recherche, UMR7589.



\begin{thebibliography}{99}


\bibitem{Ito70}
N. Itoh, {\it Prog. Theor. Phys. } {\bf  44}, 291 (1970).

\bibitem{Car73}
P. Carruthers, {\it Collectuve Phenomena} {\bf 1}, 147 (1973).

\bibitem{ILL74}
F. Iachello, W.D. Langer and W.D. Lande,
{\it Nucl. Phys.} A {\bf 219}, 612 (1974).
 
\bibitem{Col75}
J.C. Collins and M.J. Perry, 
{\it Phys. Rev. Lett.} {\bf 34}, 1353 (1975).

\bibitem{Hag84}
R. Hagedorn, in Quark Matter'84, p53, K. Kajantie, ed.,
Springer Lecture Notes in Physics, Vol. 221 (Springer, Berlin, 1985).

\bibitem{Kar00}
F. Karsch, E. Laermann and A. Peikert, 
{\it Phys. Lett.} B {\bf 478},  447 (2000).


\bibitem{abundance}
 J. Rafelski, 
``Extreme States of Nuclear Matter",
 p.\,282 in  {\it Future Relativistic Heavy Ion  Experiments}, 
R. Bock and R. Stock, Eds., GSI Report 1981-6; \\
J. Rafelski, and R. Hagedorn, 
``From Hadron Gas to Quark Matter II",
p.\,253 in {\it Statistical Mechanics of Quarks and Hadrons},
H. Satz, ed.; (North Holland, Amsterdam, 1981);\\
 J. Rafelski,  
``Hot Hadronic Matter'',
p.\,619 in {\it New Flavor and Hadron Spectroscopy},
 J. Tran Thanh Van, Ed., Ed. Fronti\`eres (Paris 1981);\\
J. Rafelski, 
``Extreme states of nuclear matter''.
{\it Nucl. Physics} A {\bf 374}, 489c (1982).

\bibitem{RM82}
{J. Rafelski and B. M\"uller}, 
{\it Phys. Rev. Lett} {\bf 48}, 1066 (1982); {\bf 56}, 2334E (1986);\\
{P. Koch, B. M\"uller and J. Rafelski},
{\it Z. Phys.} A {\bf 324}, 453 (1986).

\bibitem{Raf82}
J. Rafelski, 
{\it Phys. Rep.} {\bf 88}, 331 (1982).

\bibitem{RD83}
J. Rafelski and M. Danos, 
{\it Perspectives in High Energy Nuclear Collisions}, 
NBS-IR 83-2725 Monograph, U.S. Department of Commerce, 
National Bureau of Standards, June 1983;\\
Updated version  appeared  in 
{\it Nuclear Matter under Extreme Conditions}, D. Heiss, Ed.,
Springer Lecture Notes in Physics {\bf 231}, pp.\,362-455 (1985).

\bibitem{Koc85}
P.  Koch and J. Rafelski, {\it Nucl. Phys.} A {\bf 444}, 678 (1985).

\bibitem{KMR86}
P.~Koch, B.~M\"uller and J.~Rafelski, 
{\it Phys. Rep.} {\bf 142}, 167 (1986).


\bibitem{WA97p}
F. Antinori {\it et al.},  WA97 Collaboration 
{\it Nucl. Phys.} A {\bf 663}, 717 (2000);\\
E. Andersen {\it et al.}, WA97   collaboration,
{\it Phys. Lett.} B {\bf 433}, 209 (1998);\\
E. Andersen {\it et al.}, WA97  collaboration, 
{\it Phys. Lett.} B  {\bf 449}, 401 (1999).

\bibitem{Let00}
J. Letessier and J. Rafelski,
{\it Int. J. Mod. Phys.} E {\bf 9}, 107, (2000), and
references therein.

\bibitem{Ham00}
S. Hamieh, J. Letessier, and J. Rafelski,
{\it Phys. Rev.} C {\bf 62}, 064901 (2000).

\bibitem{Raf00}
J. Rafelski and J. Letessier, 
{\it Phys. Rev. Lett.}  {\bf 85}, 4695 (2000).

\bibitem{Tor00}
G. Torrieri and J. Rafelski,
submitted to New J. Phys., 
E-print hep-ph/0012102


\bibitem{Ant00}
F. Antinori  {\it et al.},  WA97 Collaboration 
{\it Eur. Phys. J.} C {\bf 14}, 633, (2000), and
private communication.


\bibitem{RD87}
J.~Rafelski and M.~Danos,
{\it Phys. Lett. {\rm B}}, {\bf 192}, 432 (1987).

\bibitem{Raf99} 
J. Rafelski  and J.  Letessier, 
{\it Phys. Lett.} B {\bf 469}, 12 (1999).


\bibitem{Bas00}
S. Bass {\it et al.}, 
{\it Nucl. Phys.} A {\bf 661}, 260 (2000).

\bibitem{Acta96} 
{J. Rafelski, J. Letessier and A. Tounsi},
{\it Acta Phys.\,Pol.}\,B {\bf 27}, 1035 (1996).

\bibitem{Pes00}
A.~Peshier, B.~K{\" a}mpfer and G.~Soff, 
{\it Phys. Rev.} {\rm C} {\bf 61}, 45203, (2000).

\bibitem{Vij95}
H.~Vija and M.H. Thoma, 
{\it Phys. Lett.} {\rm B} {\bf 342}, 212, (1995).

\bibitem{Chi78}
S.A. Chin, 
{\it Phys. Lett.} {\rm B}, {\bf 78}, 552, (1978).

\bibitem{entro}
{J. Letessier, A. Tounsi and J. Rafelski}, 
{\it Phys. Rev. }C {\bf 50}, 406 (1994);\\
{J. Rafelski, J. Letessier and A. Tounsi},  
{\it  Acta Phys. Pol.} A {\bf 85}, 699 (1994).


\bibitem{HR80}
R. Hagedorn and J. Rafelski
{\it Phys. Lett.} B {\bf  97}, 136 (1980).


\bibitem{Cso94}
T. Cs\"org\H{o}, and L.P. Csernai, 
{\it Phys. Lett.} B {\bf 333}, 494 (1994).

\bibitem{NA49stop}  
H.\,Appelsh\"auser {\it et al.}, NA49, 
{\it Phys. Rev. Lett.} {\bf 82}, 2471 (1999).


\bibitem{Raf91}
J. Rafelski,
{\it Phys. Lett.} B {\bf 262}, 333 (1991);\\
J. Letessier and J. Rafelski, 
{\it Acta Phys. Pol.} B {\bf 30}, 153 (1999).



\bibitem{Cse95}
L.P. Csernai, I.N. Mishustin, 
{\it Phys. Rev. Lett.} {\bf  74},  5005, (1995). 

\bibitem{Bir99}
T.S. Bir\'o, P. L\'evai, J. Zim\'anyi,
{\it Phys. Rev.} C {\bf 59}, 1574 (1999).

\bibitem{Mis99}
I.N. Mishustin,
{\it Phys. Rev. Lett.}  {\bf 82}, 4779 (1999).


\bibitem{LanHyd}
L.D. Landau and E.M. Lifschitz, ``Fluid Mechanics'',
(Oxford 1999); S. Weinberg, ``Gravitation and Cosmology'',  
(New York, 1972).


\bibitem{Kab99}
S. Kabana  {\it et al.}, NA52 collaboration,
{\it Nucl. Phys.} A {\bf 661}, 370c (1999);\\
S. Kabana  {\it et al.}, NA52 collaboration,
{\it J. Phys.} G {\it Nucl. Part. Phys.} {\bf 25}, 217 (1999).




\bibitem{WA98pi0}
M. Aggarwal {\it et al.},  WA98 Collaboration 
{\it  Phys.\,Rev.\,Lett.}\,{\bf 83}, 926 (1999); we thank T. Petizmann
for making available the data as published. 



\bibitem{NuX98}
I.G. Bearden {\it et al.},  NA44 Collaboration
{\it Phys.\,Rev.\,Lett.}\,78, 2080 (1997).


\bibitem{Fey77}
R.D. Fields and R.P. Feynman, 
{\it Phys. Rev.} D {\bf 15}, 2590 (1977); and 
R.P. Feynman, R.D. Fields and G.C. Fox,
{\it Phys. Rev.} D {\bf 18}, 3320 (1978).


\bibitem{Let93} 
J. Letessier,  {\it et al.},
{\it Phys. Rev. Lett.} {\bf 70}, 3530 (1993);
%
{\it Phys.\ Rev.} D {\bf 51}, 3408 (1995).



\bibitem{BM99}
P. Braun-Munzinger, I. Heppe and J. Stachel,
{\it Phys.\,Lett.}\,B {\bf 465}, 15 (1999).


\bibitem{Tor01}
G. Torrieri and J. Rafelski,
{\it $\Sigma^*$ as signature of freeze-out dynamics},
E-print hep-ph/01
\end{thebibliography}
\end{document}